\documentclass[a4paper,11pt]{article}
\pdfoutput=1 

\usepackage{jcappub}
\usepackage{slashed}
\usepackage{graphicx}% Include figure files
\usepackage{dcolumn}% Align table columns on decimal point
\usepackage{bm}% bold math
\usepackage[dvipsnames]{xcolor}
\def\beq{\begin{equation}\begin{aligned}}
\def\eeq{\end{aligned}\end{equation}}

\newcommand{\be}{\begin{equation}} 
\newcommand{\ee}{\end{equation}}

\newcommand{\dd}{\mathrm{d}}
\newcommand{\Msun}{\ensuremath{\mathrm{M}_\odot}}

\def\chieff{\chi_{\rm eff}}

\begin{document}

\preprint{SCIPP 19/02}

\title{Unraveling the origin of black holes from effective spin measurements with LIGO-Virgo}% Force line breaks with \\
\author[a,b]{Nicolas Fernandez} 
\author[a,b]{and Stefano Profumo}

\affiliation[a]{Department of Physics, 1156 High St., University of California Santa Cruz, Santa Cruz, CA 95064, USA}
\affiliation[b]{Santa Cruz Institute for Particle Physics, 1156 High St., Santa Cruz, CA 95064, USA}

\emailAdd{nfernan2@ucsc.edu}
\emailAdd{profumo@ucsc.edu}
%\date{\today}
% It is always \today, today,
             %  but any date may be explicitly specified

\abstract{We investigate how to use information on the effective spin parameter of binary black hole mergers from the LIGO-Virgo gravitational wave detections to discriminate the origin of the merging black holes. We calculate the expected probability distribution function for the effective spin parameter for primordial black holes. Using LIGO-Virgo observations, we then calculate odds ratios for different models for the distribution of black holes' spin magnitude and alignment. We evaluate the posterior probability density for a possible mixture of astrophysical and primordial black holes as emerging from current data, and calculate the number of future merger events needed to discriminate different spin and alignment models at a given level of statistical significance.}


\maketitle
\flushbottom

%\tableofcontents

%%%%%%%%%%%%%%%%%%%%%%%%%%%%%%%%%%%%%%%%%%%%%%%%%%%%%%%%%%%%%%%%%%%%%%%%%%%%%%%%%%%%%%%%%
%%%%%%%%%%%%%%%%%%%%%%%%%%%%%%%%%%%%%%%%%%%%%%%%%%%%%%%%%%%%%%%%%%%%%%%%%%%%%%%%%%%%%%%%%
%%%%%%%%%%%%%%%%%    Introduction 
%%%%%%%%%%%%%%%%%%%%%%%%%%%%%%%%%%%%%%%%%%%%%%%%%%%%%%%%%%%%%%%%%%%%%%%%%%%%%%%%%%%%%%%%%
%%%%%%%%%%%%%%%%%%%%%%%%%%%%%%%%%%%%%%%%%%%%%%%%%%%%%%%%%%%%%%%%%%%%%%%%%%%%%%%%%%%%%%%%%
\section{Introduction}

After the detection of binary black hole (BBH) merger events with LIGO-Virgo \cite{Abbott:2016blz, LIGOScientific:2018jsj}, the question of the {\em physical origin} of the black holes has become somewhat pressing. In particular, there has been some significant interest in the possibility that some or all of the BBH events originate from {\em primordial} black holes (PBH) \cite{Bird:2016dcv, Blinnikov:2016bxu}, i.e., black holes originating from large over-densities in the very early universe rather than from the collapse of stellar objects (see e.g. \cite{Carr:2009jm, Carr:2016drx}). Interestingly, this interpretation is compatible with the notion that these PBH could be the {\em dark matter} needed for a consistent picture of the early and large-scale structure of the universe (see e.g. \cite{Bird:2016dcv}). Whether or not this possibility is ruled out is the focus of intense debate, the key issues at stake including (1) the problem of matter accretion that might produce significant-enough accelerated charged particles to perturb in a measurable, and excessive way the cosmic microwave background photons \cite{Poulin:2017bwe, Nakama:2017xvq} (see however \cite{Ali-Haimoud:2016mbv}); (2) the problem of disruption of small-scale structure by the relatively massive black holes that would make up the dark matter \cite{Brandt:2016aco, Koushiappas:2017chw} (see however \cite{Li:2016utv}); and (3) limits from gravitational lensing of type Ia supernovae  \cite{Zumalacarregui:2017qqd} (see however \cite{Garcia-Bellido:2017imq}).

These constraints notwithstanding, it is known (and it is reviewed below) that PBH that were produced during a radiation-dominated cosmological epoch have low intrinsic spin magnitude. As a result, a generic prediction of the picture where the LIGO-Virgo BBH events are in part or all PBH mergers is that the effective spin parameter (to be defined below, and which depends among other things on the intrinsic black hole spin magnitudes) is very low. Incidentally, it is important to point out that this fact (that binary mergers of PBH should have a low effective spin parameter) does depend on cosmology: it has been shown, for instance, that if the universe went through a matter-domination phase during PBH formation, then in fact the intrinsic spin of the resulting PBHs is generically {\em close to maximal} \cite{Harada:2017fjm}. Furthermore, other possibilities might arise for cosmologies where the early universe was neither matter- nor radiation-dominated at early times, as we explored recently \cite{DEramo:2017gpl, DEramo:2017ecx}. With this caveat in mind, we shall hereafter assume radiation domination at the time of PBH formation.

The key motivation for the present study is that, so far, 9 out of 10 of the LIGO-Virgo BBH mergers are compatible with very low effective spin parameters. It seems timely, therefore, to assess what the {\em predicted} probability density for PBH's effective spin parameter is, and to compare it with current observations; additionally, we intend to explore {\em how many events} it will take to differentiate in a statistically robust fashion different effective spin parameter models. Although at present there is no firm prediction for astrophysical black holes' effective spin parameter, and there is debate concerning selection and bias effects in the detected BBH events, our study intends to point out that (i) current observations of the effective spin parameter are largely compatible with a dominant PBH component in the BBH mergers, and that (ii) in the future, the effective spin parameter distribution could help discriminating PBH from ``ordinary'', astrophysical black holes.

Aspects of the question we address here have been considered in the recent literature, with different assumptions and methods. Soon after the detection of the first four BBH mergers \cite{Abbott:2016blz,Abbott:2017vtc, Abbott:2017gyy,Abbott:2017oio,TheLIGOScientific:2016pea}, ref.~\cite{Farr:2017uvj} argued that information on the effective spin could be used to distinguish between aligned versus isotropic angular distributions; specifically, \cite{Farr:2017uvj} concluded that as long as the black hole spin values are not intrinsically small (which, however, might well be a distinct possibility, as pointed out e.g. by ref.~\cite{Belczynski:2017gds}) then an aligned angular distribution is strongly disfavored. Additionally, ref.~\cite{Farr:2017uvj} also showed how with relatively few additional events, the odds ratio would conclusively point in one direction or another (i.e., isotropic or aligned).

Ref.~\cite{Vitale:2017cfs} reiterated how scenarios considered in the formation of astrophysical black hole binaries naturally lead to isotropic (for dynamical capture) or near-aligned (for common envelope evolution) black hole spins. They also showed that Bayesian statistics allows one to distinguish, at a given confidence level, which {\em fraction} of the binaries are preferentially aligned versus isotropically distributed. A similar analysis was conducted in ref.~\cite{Farr:2017gtv}, with the additional points that a discrimination between isotropic and aligned spin distributions might be possible even regardless of the intrinsic spin magnitude distribution. They also showed that once an aligned or isotropic spin distribution is established, it is possible to re-construct the spin magnitude distribution with a high degree of confidence. The possibility of disentangling the existence of sub-populations of binary black holes with different spin orientations was explored in ref.~\cite{Vitale:2015tea, Stevenson:2017dlk}, which also pointed out how a ``pure'' distribution would be statistically preferred with relatively few events (see also ref.~\cite{Wysocki:2018mpo}).

In this study, we utilize the intrinsic spin distribution of PBHs from the results of ref.~\cite{Chiba:2017rvs} to calculate (to our knowledge for the first time) the predicted prior probability distribution for the  effective spin parameter. Ref.~\cite{Chiba:2017rvs} assumes that there is no correlation between the overdensity leading to the formation of the PBH and its spin, and that the probability density for the spin distribution as a function of the overdensity is flat. A recent study, ref.~\cite{DeLuca:2019buf}, challenges these assumptions, and finds a peaked distribution for the PBH spin parameter, which critically depends on the width of the power spectrum peak giving rise to the PBH, and on the relative abundance of PBH. As a result, the spin distribution is significantly narrower than what was predicted in ref.~\cite{Chiba:2017rvs} (see also refs.~\cite{Mirbabayi:2019uph, He:2019cdb}). In what follows, we compare the resulting spin distribution for a variety of assumptions for PBH formation as outlined in ref.~\cite{DeLuca:2019buf}, and compare it to our benchmark choice which reflects the results of ref.~\cite{Chiba:2017rvs} (and which, as we explain below, can be seen as a limiting case of the setup of ref.~\cite{DeLuca:2019buf}).

To these ends, in this study we first explore, in sec.~\ref{sec:spin_distribution}, the theoretical prediction for the effective spin parameter distribution for PBH. We outline the assumed astrophysical black holes spin magnitude distribution we consider, and we review LIGO-Virgo observations; in sec.~\ref{sec:methods_results} we compare the odds ratios for the models we consider and study the favored ``mixture'' of different such models; we then forecast how future events will inform both the odds ratios and the inference of the relative level of mixing of different models. Finally, in sec.~\ref{sec:conclusions} we discuss our results and present our conclusions.

%%%%%%%%%%%%%%%%%%%%%%%%%%%%%%%%%%%%%%%%%%%%%%%%%%%%%%%%%%%%%%%%%%%%%%%%%%%%%%%%%%%%%%%%%
%%%%%%%%%%%%%%%%%%%%%%%%%%%%%%%%%%%%%%%%%%%%%%%%%%%%%%%%%%%%%%%%%%%%%%%%%%%%%%%%%%%%%%%%%
%%%%%%%%%%%%%%%%%    Spin Distribution 
%%%%%%%%%%%%%%%%%%%%%%%%%%%%%%%%%%%%%%%%%%%%%%%%%%%%%%%%%%%%%%%%%%%%%%%%%%%%%%%%%%%%%%%%%
%%%%%%%%%%%%%%%%%%%%%%%%%%%%%%%%%%%%%%%%%%%%%%%%%%%%%%%%%%%%%%%%%%%%%%%%%%%%%%%%%%%%%%%%%

\section{Effective Spin Distribution}
\label{sec:spin_distribution}

The spin magnitude of a Kerr black hole (BH) is commonly defined via a dimensionless spin parameter $\chi$, 
\beq
\chi = \dfrac{|\vec{S}|}{G m^{2}},
\label{Eq:chidef}
\eeq
where $\vec{S}$ and $m$ are the spin and mass of the BH, respectively. One of the most important parameters that LIGO can infer from the gravitational waveform is the {\em effective spin parameter} $\chieff$, defined as:
\beq
\label{eq:chieff}
\chieff=\dfrac{m_{1} \chi_{1} \cos \theta_{1} + m_{2} \chi_{2} \cos \theta_{2}}{m_1 + m_2},
\eeq
where $\theta_{i} = \cos^{-1} (\vec{L} \cdot \vec{S_{i}})$ is the tilt angle between the spin $\vec S_i$ and the orbital angular momentum vector $\vec{L}$. As apparent from its definition, the parameter $\chieff$ is a quantity sensitive to both the spin alignment of the two black holes with their orbit (angular momentum of the binary) before the merger, and to the magnitude of the individual spins. $\chieff$ is a dimensionless number ranging from $-1$ to $1$, where for $\chieff=1$ the spins of both black holes are perfectly aligned with their orbit, and $\chieff=-1$ the spins are perfectly anti-aligned. Values of $\chieff \approx 0$ can stem from one or both of the following physical situations: (i) the black hole intrinsic spins are anti-aligned with each other, or (ii) the magnitude of the intrinsic effective spin parameters, $\chi_{i} \ll 1$. There could be a third possibility, that both spins are perpendicular to the orbit, but this is somewhat less likely and less physically motivated.

First, let us gain some intuition about the effective spin distribution for a few binary black hole (BBH) formation channels. One possible formation channel for BBH is from massive isolated binaries through common envelope evolution, where the intrinsic spin is generally aligned along the same direction as the orbital angular momentum, meaning that  $\chieff \approx 1$. On the other hand, there exist dynamical scenarios where we expect most BBHs to have spins largely uncorrelated  with their orbit meaning that  $\chieff \approx 0$. This is the case, for instance, for BBHs formed dynamically in dense stellar environments, and it is also the case for PBHs, which additionally are predicted to have small intrinsic spins.  It is important to notice that a key consequence, and possible signature, of any isotropic formation mechanism is that the distribution of $\chieff$ is {\em symmetric around zero}, regardless of the spin magnitude distribution \cite{Farr:2017gtv}.

%%%%%%%%%%%%%%%%%%%%%%%%%%%%%%%%%%%%%%%%%%%%%%%%%%%%%%%%%%%%%%%%%%%%%%%%%%%%%%%%%%%%%%%%%
%LIGO-Virgo effective spin measurements
%%%%%%%%%%%%%%%%%%%%%%%%%%%%%%%%%%%%%%%%%%%%%%%%%%%%%%%%%%%%%%%%%%%%%%%%%%%%%%%%%%%%%%%%%
\subsection{LIGO-Virgo effective spin measurements}

\begin{table}[ht]
  \centering
\scalebox{0.9}{
\begin{tabular}{ l r r r r r r}
\hline 
Event & & \multicolumn{1}{c}{$m_1 [\Msun]$}    & &   \multicolumn{1}{c}{$m_2 \left[ \Msun \right]$}  & & \multicolumn{1}{c}{$\chieff$} \\  [0.2ex]  \hline   \hline  \\ [-1.4ex]
GW150914  & & $35.6^{+ 4.8}_{-3.0}$   & & $30.6^{+3.0}_{-4.4}$  & & $-0.01^{+0.12}_{-0.13}$ \\ [0.14cm]
GW151012  & & $23.3^{+ 14.0}_{-5.5}$  & & $13.6^{+4.1}_{-4.8}$  & &  $0.04^{+0.28}_{-0.19}$ \\ [0.14cm]
GW151226  & & $13.7^{+ 8.8}_{-3.2}$   & & $7.7^{+2.2}_{-2.6}$   & &  $0.18^{+0.20}_{-0.12}$ \\ [0.14cm]
GW170104  & & $31.0^{+ 7.2}_{-5.6}$   & & $20.1^{+4.9}_{-4.5}$  & &  $-0.04^{+0.17}_{-0.20}$ \\ [0.14cm]
GW170608  & & $10.9^{+5.3}_{-1.7}$    & & $7.6^{+1.3}_{-2.1}$   & &  $0.03^{+0.19}_{-0.07}$ \\ [0.14cm]
GW170729  & & $50.6^{+16.6}_{-10.2}$  & & $34.3^{+9.1}_{-10.1}$ & &  $0.36^{+0.21}_{-0.25}$ \\ [0.14cm]
GW170809  & & $35.2^{+8.3}_{-6.0}$    & & $23.8^{+5.2}_{-5.1}$  & &  $0.07^{+0.16}_{-0.16}$ \\ [0.14cm]
GW170814  & & $30.7^{+ 5.7}_{-3.0}$   & & $25.3^{+2.9}_{-4.1}$  & &  $0.07^{+0.12}_{-0.11}$ \\ [0.14cm]
GW170818  & & $35.5^{+7.5}_{-4.7}$    & & $26.8^{+4.3}_{-5.2}$  & &  $-0.09^{+0.18}_{-0.21}$ \\ [0.14cm]
GW170823  & & $39.6^{+ 10.0}_{-6.6}$ & & $29.4^{+6.3}_{-7.1}$  & &  $0.08^{+0.20}_{-0.22}$ \\ [1ex] \hline
\end{tabular}
}
\caption {Selected parameters of the ten BBH mergers events detected during LIGO's O1 and O2 runs. The parameters are median values, with 90\% credible intervals \cite{LIGOScientific:2018mvr}.}
\label{table: parameters}
\end{table}

We list in Table \ref{table: parameters} the relevant observed properties of the 10 BBH merger events we consider in our study: the masses of each individual black hole $m_1$ and $m_2$ (columns 2 and 3), and the corresponding dimensionless effective spin $\chieff$ (column 4). It is worth pointing out that before the first gravitational wave detection, LIGO was expecting 33-100 more NS-NS binary events compared with BH-BH binaries \cite{Abadie:2010cf} (see however \cite{Lipunov:1996nj, Lipunov:1997xm} for different earlier predictions closer to what currently observed); however, LIGO's O1 and O2 run showed that the rate of BH-BH binaries is an order of magnitude {\em greater} than the NS-NS binaries. Furthermore, the range of  black hole masses was expected to be from  $\sim 5 \Msun$ to $\sim 15 \Msun$ \cite{Mandel:2009nx}. As a result, the first detection (GW150914) came somewhat as a surprise, because previously-known black holes were significantly lighter than the inferred masses, among various reasons. 

As evident in Table \ref{table: parameters}, the majority of black holes are over $25 \Msun$ with the heaviest being $50 \Msun$ (GW170729).  The nature of this new population of heavy stellar-mass black holes is still debated in the literature \cite{Spera:2017fyx, Kimball:2019mfs,Kruckow:2018slo,Fishbach:2017dwv}; notice that a reason for observing more massive binary mergers over lighter ones could partly be explained as a selection bias \cite{Vitale:2016aso}, since more massive BBH mergers produce a louder signal, and therefore the accessible space-time volume is larger than for lighter systems.
It has been proposed that there exists a mass gap ($\sim 50 \Msun - 150\Msun$) due to pair{-}instability supernovae for stellar black holes \cite{Heger:2001cd, Belczynski:2016jno,Spera:2017fyx} and therefore, that the black hole masses cannot be arbitrary large. There have also been claims of a cutoff at high masses in the current detections made by LIGO to date \cite{Fishbach:2017zga,Talbot:2017yur,Roulet:2018jbe,Bai:2018shq}.

\begin{figure}[ht]
\center\includegraphics[width=1
\textwidth]{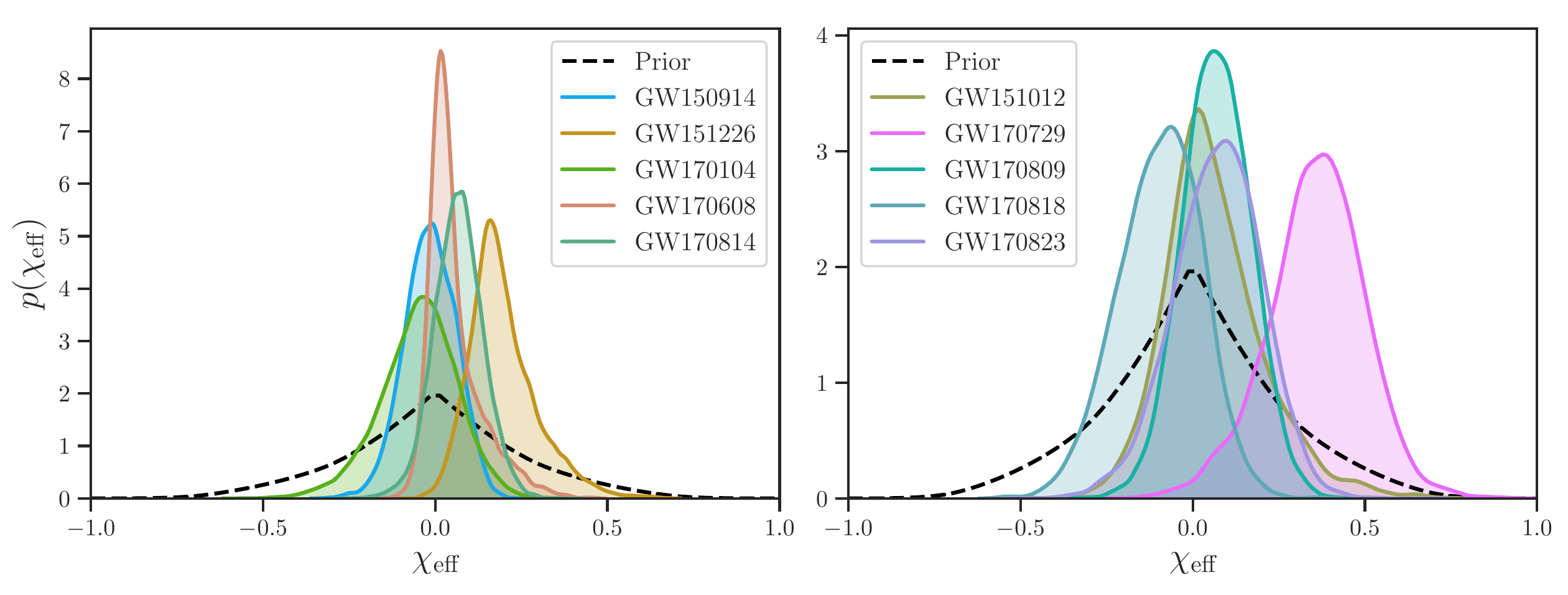}
\caption{Posterior probability densities for the effective aligned spin magnitude $\chieff$ for the 10 events from the LIGO-Virgo observations \cite{LIGOScientific:2018mvr} as given in the files downloadable at \href{ https://dcc.ligo.org/LIGO-P1800370/public}{https://dcc.ligo.org/LIGO-P1800370/public}. Notice the difference in the vertical scale for the left and right panels.}
\label{fig:Chi_eff}
\end{figure}

The last column of Table \ref{table: parameters} shows the most interesting parameter for this work: the effective spin $\chieff$, defined by Eq.~(\ref{eq:chieff}). The listed 10 observed events appear to disfavor high spin magnitude aligned with the orbital
angular momentum, unlike the large spin values (near to the maximum possible value) observed in the majority of black holes in X-ray binaries \cite{McClintock:2013vwa,Gou:2011nq}. Most events are consistent with $\chieff=0$, with two exceptions: GW170729 and GW151226. These two events show evidence of positive, but  relatively small, $\chieff$ values.
Fig.~\ref{fig:Chi_eff} shows the posterior distribution of $\chieff$ for the ten events observed by LIGO's O1 and O2 run and the prior assumed by the LIGO Collaboration\footnote{\href{https://dcc.ligo.org/LIGO-P1800370/public}{https://dcc.ligo.org/LIGO-P1800370/public}}.

It is important to note that the LIGO Collaboration used Bayesian statistics to analyze the data and to infer the source properties of all ten BBH gravitational wave events \cite{LIGOScientific:2018mvr}. This means that one needs to properly define prior probability density distributions. While ideally the conclusions should be robust and fairly independent under the choice of different priors, if the data are only mildly informative, priors could influence the statistical inference on the source parameters (see ref. \cite{Thrane:2018qnx} for a discussion on this point). An analysis of the importance and effect of the choice of priors on the first three LIGO events has been carried out in ref.~\cite{Vitale:2017cfs}. One should also bear in mind that there certainly exist selection bias effects for the posterior distribution of $\chieff$. For instance, sources with positive $\chieff > 0$ have a more clear signal, due to longer time orbiting before merging, therefore allowing to better constrain the waveform (see e.g. refs.~\cite{Ng:2018neg,Campanelli:2006uy}).

%%%%%%%%%%%%%%%%%%%%%%%%%%%%%%%%%%%%%%%%%%%%%%%%%%%%%%%%%%%%%%%%%%%%%%%%%%%%%%%%%%%%%%%%%
%PBH Spin Distribution
%%%%%%%%%%%%%%%%%%%%%%%%%%%%%%%%%%%%%%%%%%%%%%%%%%%%%%%%%%%%%%%%%%%%%%%%%%%%%%%%%%%%%%%%%
\subsection{PBH spin distribution}
\label{sec:PBH Spin Distribution}

Given a spin distribution for the intrinsic spin of individual primordial black holes, and the assumption of isotropy in the spin-orbit alignment, one can calculate the distribution of $\chieff$. We follow here ref.~\cite{Chiba:2017rvs} in assuming that the  distribution function for the intrinsic spin magnitude of a single PBH can be closely approximated by the Gaussian functional form
\beq
p(\chi) \approx  \exp{\left[- \dfrac{\chi^{2}}{2\sigma^{2}}\right]}. 
\eeq
The parameter $\sigma$ is, in principle, calculable given the spectrum of density perturbations leading to PBH formation in the early universe. Absent this, and in view of the fact that there might be circumstantial evidence for a PBH origin of at least some of the detected BBH merger events \cite{Clesse:2017bsw}, here we infer the value of $\sigma$ directly from observations. To this end, we define a probability distribution for $\chi$ as a half Gaussian 
\beq
\label{Eq:PBH_Probability_chi}
p(\chi;\mu,\sigma)=\frac{\mathcal{N(\mu,\sigma)}}{\sqrt{2 \pi} \sigma} \exp \left[ -\frac{(\chi-\mu)^{2}}{2 \sigma^{2}} \right], 
\eeq
entertaining a possible non-zero value for $\mu$. In Eq.~(\ref{Eq:PBH_Probability_chi}) $\mathcal{N(\mu,\sigma)}$ is the appropriate normalization constant. Notice that the function is only defined in the interval $\chi \in [0,1]$ and zero otherwise (see the definition of $\chi$ in Eq.(\ref{Eq:chidef})). Our goal is to investigate the probability distribution of $\chieff$ resulting from an isotropic spin orientation distribution and the intrinsic spin distribution of Eq.~(\ref{Eq:PBH_Probability_chi}) and to infer the posterior probability density for the parameters $\mu$ and $\sigma$ from the LIGO-Virgo data.

We utilize here a {\em hierarchical Bayesian analysis}: we assume that the individual BHs in the binary are coming from a primordial population which is described by some hyper-parameters $\Lambda$. Then, we use the 10 LIGO{-}Virgo events to derive the posterior distribution for $\Lambda$. Our approach is analogous to refs. \cite{Farr:2017uvj,Farr:2017gtv,Talbot:2017yur,Tiwari:2018qch}. Our goal is thus to find $p\left( \Lambda  \mid  d \right)$, the probability distribution of the parameters $\Lambda$ given the data $d$, where $\Lambda$ simply represents here the parameters that describe the PBH population $\Lambda= \left\{ \mu, \sigma \right\}$. 

Assuming that the events are independent of each other, we can combine the individual likelihoods to build a joint likelihood
\beq
p\left( \{d^{i}\}  \mid  \Lambda \right) = \prod_{i=1}^{N_\mathrm{obs}}  p\left( d^{i}  \mid  \Lambda \right),
\eeq
where
\beq
p\left( d^{i}  \mid   \Lambda \right) = \int \dd \chieff^i \, p \left(  d^{i}  \mid  \chieff^i \right) p \left( \chieff^i  \mid  \Lambda \right)
\eeq
is the likelihood function for $i$th event and $p \left(  d \mid  \chieff \right)$ is the marginal likelihood, meaning that it has been marginalized over all parameters but $\chieff$. Since the LIGO-Virgo collaboration have not yet released the marginalized or full likelihoods to the general public, but have rather provided the posterior distributions which we show in Fig.~\ref{fig:Chi_eff}, we need to re{-}weight the posterior distribution of $\chieff$ to obtain the likelihoods. The last term in the integral, $p \left( \chieff  \mid  \Lambda \right)$, is the probability of measuring $\chieff$ given the parameters of our model $\Lambda$. In our case this distribution of $\chieff$ has been derived in App.~\ref{app:PBH} and the result is given in Eq.~(\ref{Eq:Prior_Chieff}). Finally, using Bayes' theorem we obtain the posterior distribution of the parameters $\Lambda$ as 
\beq
p\left( \Lambda \mid \left\{ d^i \right\} \right) \propto \left[ \prod_{i=1}^{N_\mathrm{obs}} \int \dd\chieff^i \,  p\left( d^i \mid \chieff^i \right) p\left( \chieff^i \mid \Lambda \right) \right] p\left(\Lambda \right) \,.
\eeq
Here, $p\left(\Lambda \right)$ is the prior choice for the parameters $\Lambda$, which we  take here to be uninformative (i.e. we use a flat prior for both $\mu$ and $\sigma$).  

\begin{figure}
    \centering
    \includegraphics[width= .48\textwidth]{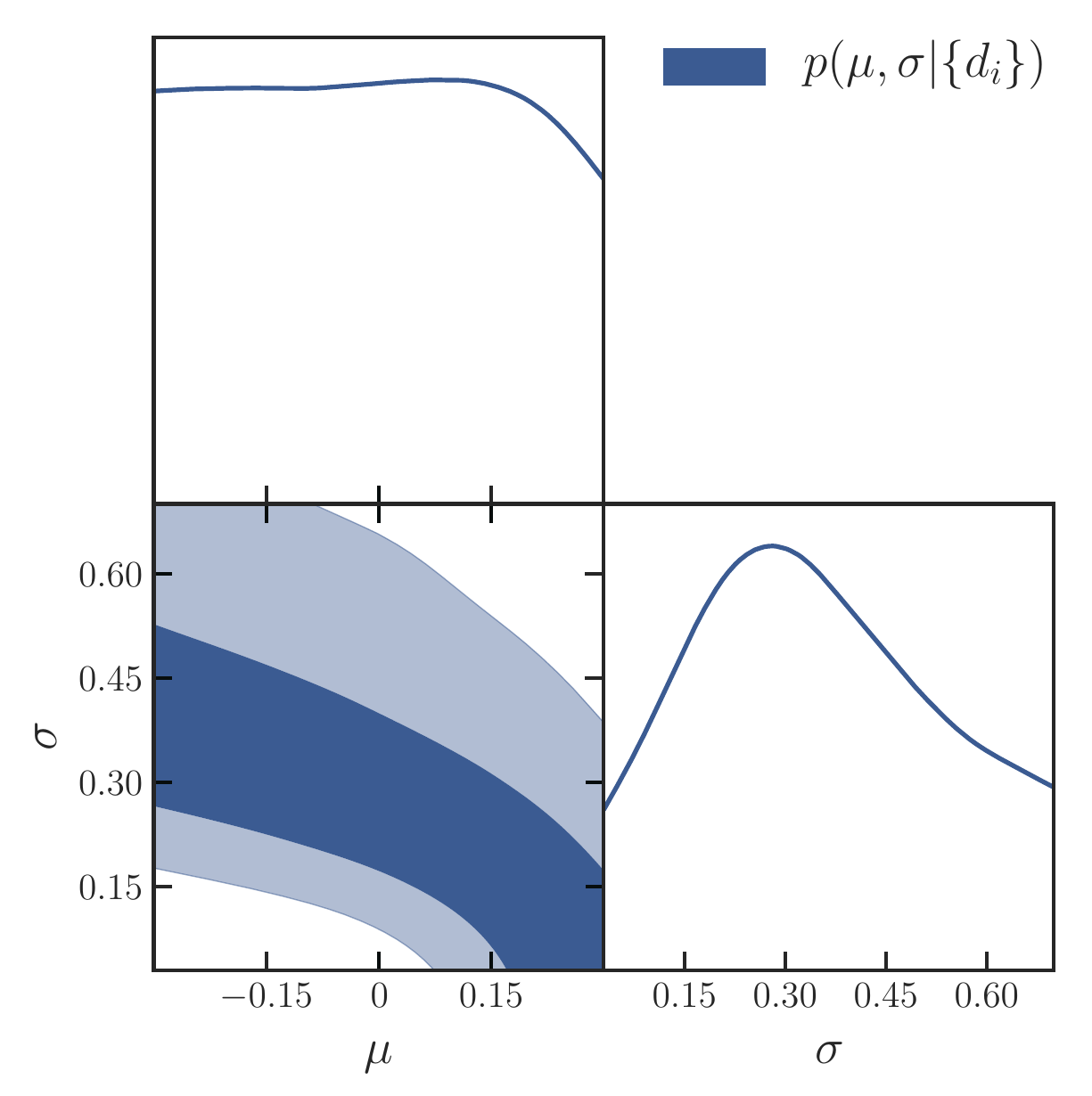}
    \includegraphics[width= .48\textwidth]{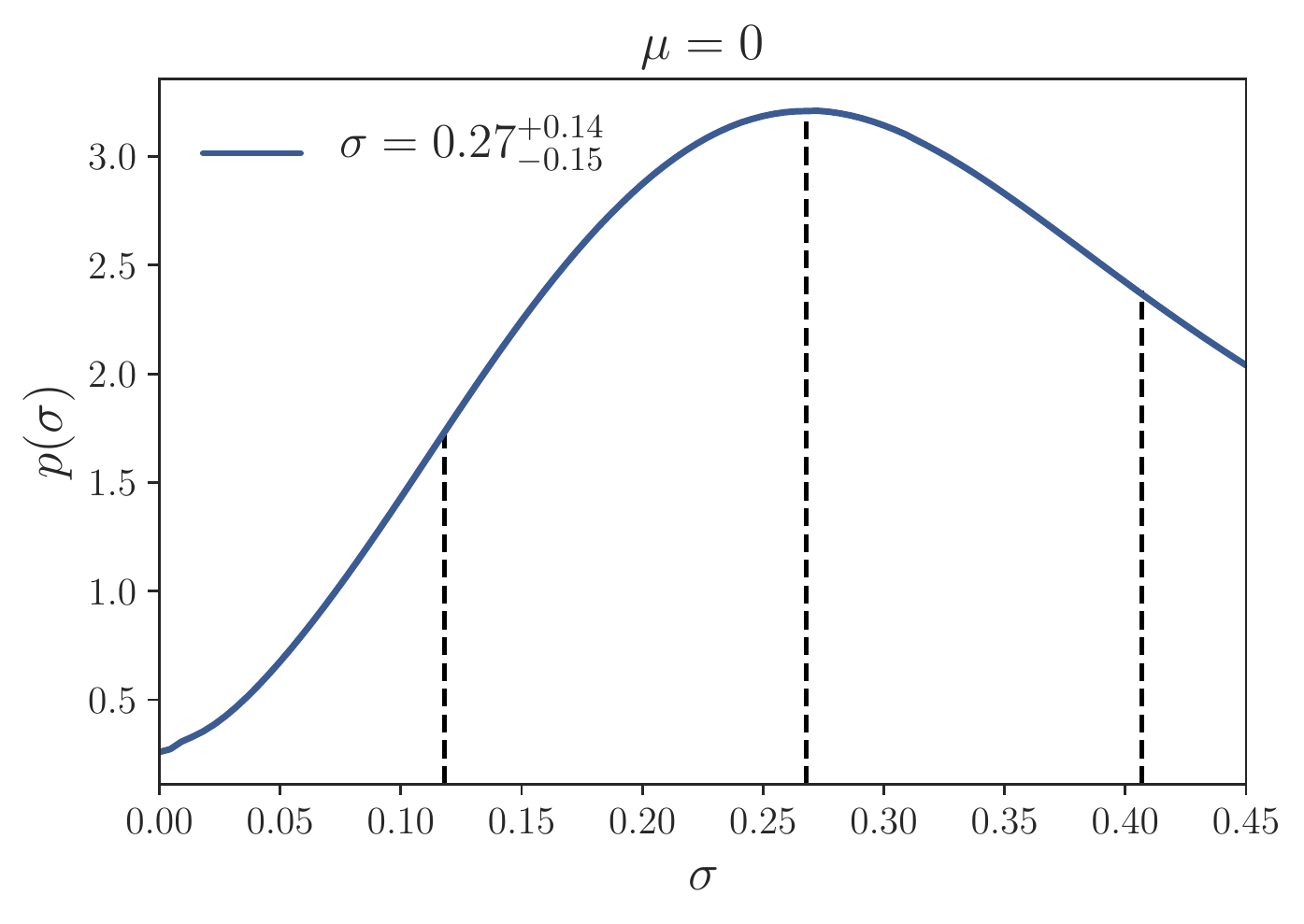}
    \caption{\textit{Left}: Marginalized probability density functions for the $\mu$ and $\sigma$ parameters describing the intrinsic PBH spin magnitude distribution. Colored contours show the 50\% and 90\% credible intervals. \textit{Right}: Probability density functions for $\sigma$ parameters with $\mu=0$ with 90\% credible intervals.}
    \label{fig:sigma_mu}
\end{figure}

The posterior distribution for the hyper-parameters $\mu$ and $\sigma$ describing the putative PBH population given the 10 LIGO-Virgo events is shown in the left panel in Fig.~\ref{fig:sigma_mu}. The distribution for $\mu$ is almost flat, except for $\mu>0.15$, a slightly disfavored range of values. In contrast, the distribution for $\sigma$ offers more information: one can clearly see a peak around $\sim 0.3$. Notice that $\mu$ and $\sigma$ are anti-correlated, as expected: given a half Gaussian with zero mean and fixed width, one can find an approximately equivalent distribution with a {\em negative} mean and larger width; conversely, a distribution with a positive mean will correspond to one with mean zero and a narrower width. 

Finally, in the right panel of Fig.~\ref{fig:sigma_mu} we show the posterior distribution of $\sigma$ when we set $\mu=0$. This choice is motivated by the analytical findings of ref.~\cite{Chiba:2017rvs} (see their Eq.(17) and figure 2), and is approximately valid even in the scenario discussed in ref.~\cite{DeLuca:2019buf}, where the preferred value of $\mu$ is non-zero but extremely small (see their Eq.(8.7) and Fig.7).

Based on the 10 LIGO-Virgo events under consideration here, the best fit value for $\sigma$ is $0.27^{+0.14}_{-0.15}$, which, remarkably, is {\em inconsistent} with zero at 90\% confidence level. Notice that the distribution does not change much compared to the general case where we marginalize over $\mu$, indicating that the data is largely insensitive to the value of $\mu$.

%%%%%%%%%%%%%%%%%%%%%%%%%%%%%%%%%%%%%%%%%%%%%%%%%%%%%%%%%%%%%%%%%%%%%%%%%%%%%%%%%%%%%%%%%
%Benchmark Spin Models for Astrophysical BH
%%%%%%%%%%%%%%%%%%%%%%%%%%%%%%%%%%%%%%%%%%%%%%%%%%%%%%%%%%%%%%%%%%%%%%%%%%%%%%%%%%%%%%%%%
\subsection{Benchmark spin models for astrophysical BH}

We are interested in comparing the  $\chieff$ distribution for PBH, discussed above, with what predicted in the case of astrophysical black holes. Given the current status of observations and the output of population synthesis codes, it is presently unwarranted to try to reproduce specific binary black hole spin distribution reflective of given astrophysical formation processes. Rather, it has become somewhat customary in the literature to adopt simplified benchmark models for the alignment and intrinsic spin distributions of astrophysical black holes, following what proposed in ref.~\cite{Farr:2017gtv}, and endorsed and utilized by the LIGO collaboration \cite{LIGOScientific:2018jsj} and by others (see e.g. \cite{Tiwari:2018qch}). We shall assume that the merging black holes have equal mass (see a discussion of the effect of unequal mass mergers on $\chieff$ in the Appendix), and that the distribution for the spin magnitude is statistically independent of that for the spin alignment.

Noting that the spin directions for isolated binary black holes are thought to be dominantly aligned (see e.g. \cite{1993MNRAS.260..675T, Miller:2014aaa,Gerosa:2018wbw}), we choose a distribution for the spin direction with perfect alignment as an extreme case. We note that this assumptions reflects under any circumstances an extreme, idealized situation for a variety of reasons: for instance, there exists evidence for spin-orbit {\em misalignment} in black hole X-ray binaries \cite{2010ApJ...719L..79F}, and effects from the supernova explosion could also contribute to tilt the spin-orbit angle (natal kicks) \cite{Wysocki:2017isg}. We study the systematic effects of relaxing the assumption of perfect alignment in the Appendix, see in particular Fig.~\ref{fig:chieff_Priors_sigma_q_35}, bottom panel.

\begin{figure}
    \centering
    \includegraphics[width= .48\textwidth]{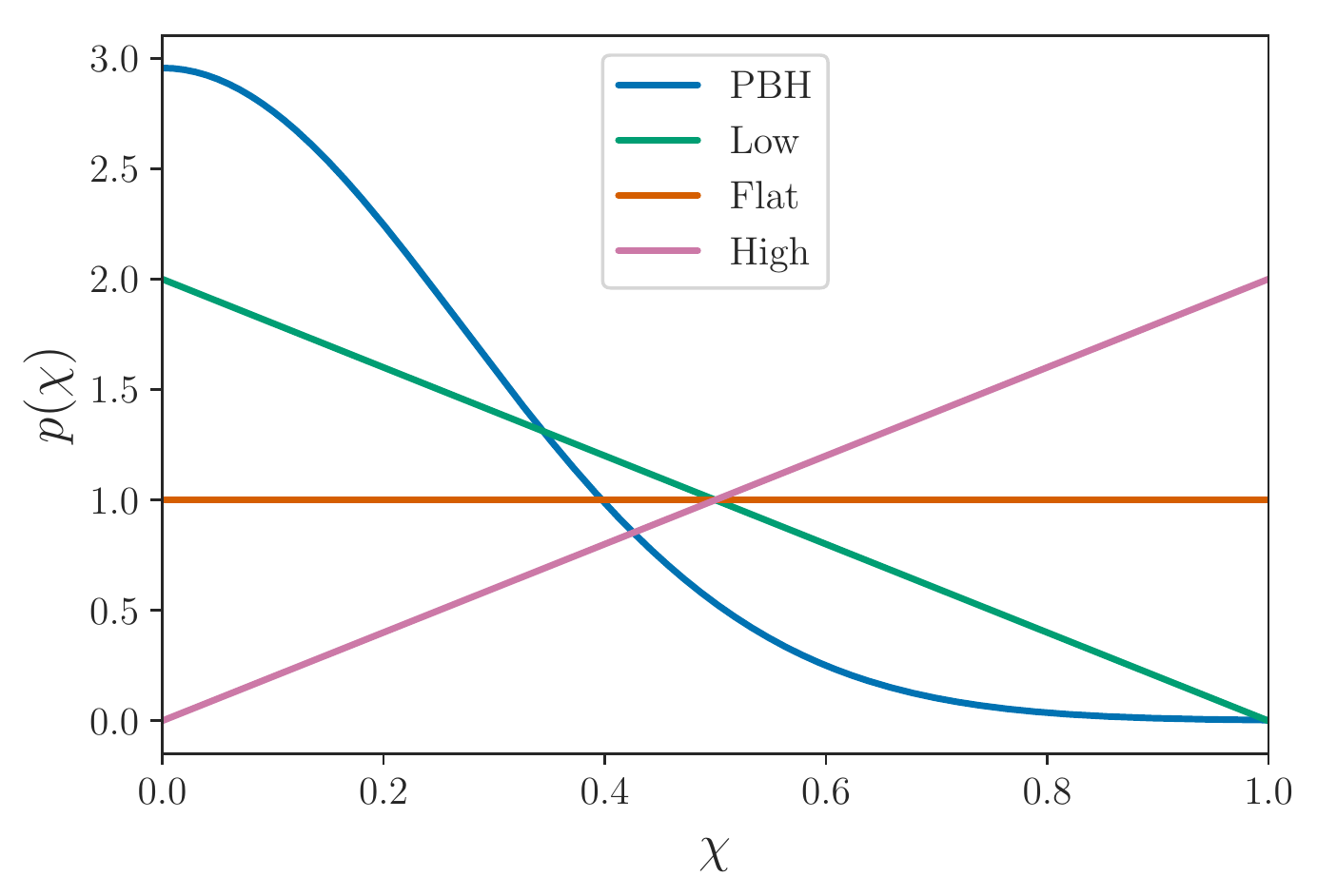}
    \includegraphics[width= .48\textwidth]{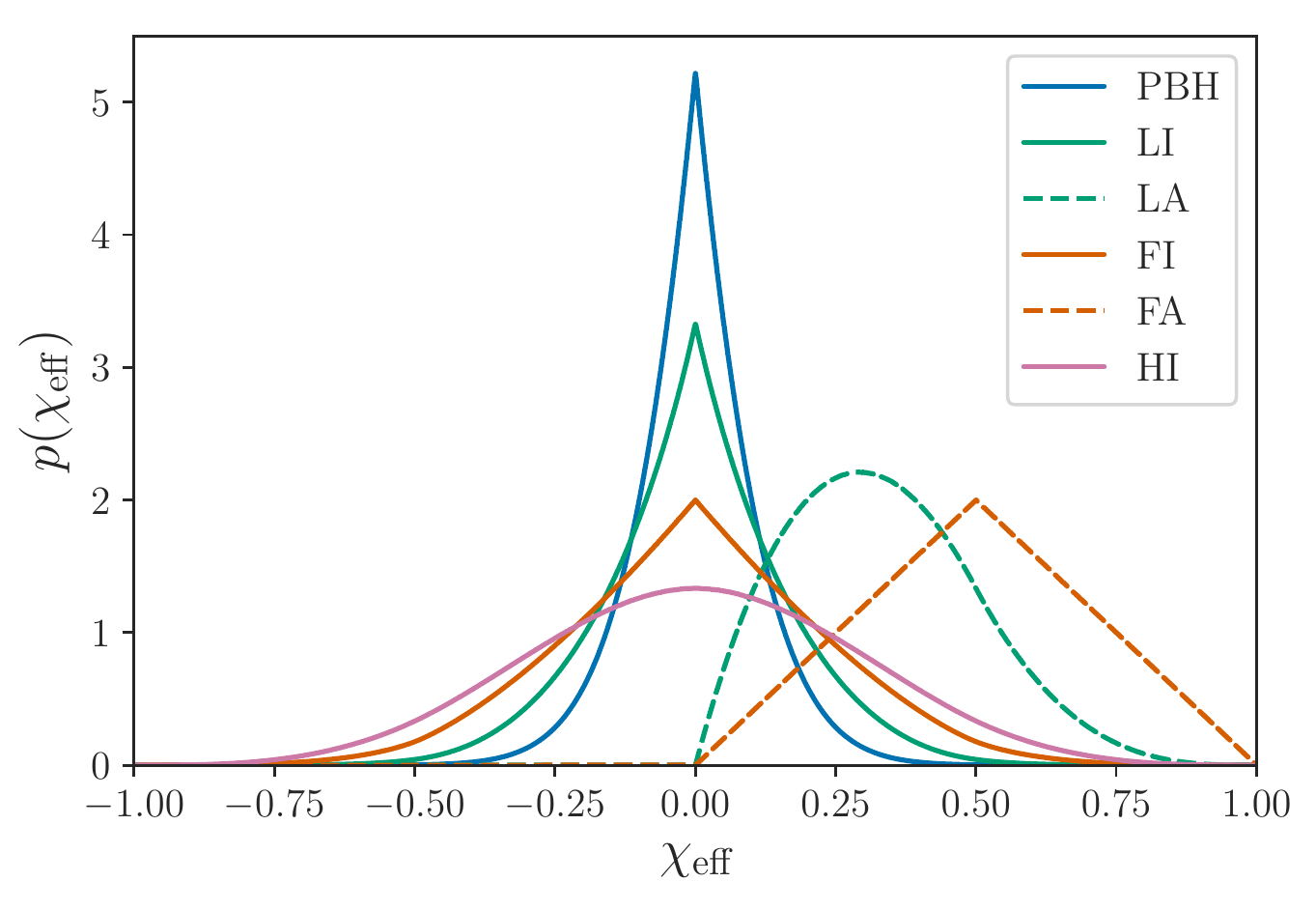}
    \caption{\textit{Left}: Normalized spin magnitude distributions for PBH, low, flat and high spin models. \textit{Right}: Prior distributions for $\chieff$ for the different models under consideration here. Solid lines indicate isotropic models while the two dashed lines to the far right (peaking at $\chi_{\rm eff}\ne0$) correspond to spin-orbit aligned ones.}
    \label{fig:Priors}
\end{figure}

We parameterize the astrophysical spin magnitude following the spin distribution proposed by \cite{Farr:2017gtv} and used by LIGO \cite{LIGOScientific:2018jsj} and \cite{Tiwari:2018qch}. The models consist of $3$ different spin magnitude distributions:
\begin{itemize}
    \item a {\em low} (intrinsic) spin distribution $p(\chi)=2(1-\chi)$ (L),
    \item a {\em flat} spin distribution $p(\chi)=1$ (F), and 
    \item a {\em high} spin distribution $p(\chi)=2\chi$ (H).
\end{itemize} We reproduce the distributions in the left panel of Fig.~\ref{fig:Priors}, together with the PBH intrinsic spin distribution for PBH for the central value of $\sigma=0.27$ and $\mu=0$ inferred above.

Also following ref.~\cite{Farr:2017gtv}, we consider two spin-orbit distribution orientations: {\em aligned} and {\em isotropic}. Notice that the tilt angle is an excellent tracer of BBH formation channels \cite{TheLIGOScientific:2016htt}, with the aligned distribution expected for isolated binary formation channel \cite{Belczynski:2016jno,Stevenson:2017dlk,Mandel:2009nx,Marchant:2016wow}, under the simplifying assumption that the binaries remain perfectly aligned throughout their evolution (an assumption that could be violated by effects such as supernova natal kicks, although mass transfer and tidal interactions might work in the opposite direction and tend to re-align the binary). The isotropic distribution is motivated by dynamical formation mechanisms in dense stellar environments or similarly disordered assembly scenarios \cite{PortegiesZwart:1999nm, Rodriguez:2015oxa, Stone:2016wzz}, as well as by what expected for PBH \cite{Sigurdsson:1993zrm}. 

The final prior distribution for $\chieff$ for the various models under consideration is shown in the right panel of Fig.~\ref{fig:Priors}, where we do not include the HA model which is already strongly excluded by data. Notice that since the intrinsic spin distribution is positive-definite, ``aligned'' models do not allow for negative values of $\chieff$; finally, also notice how the prior distribution for ``isotropic'' models is symmetric in $\chieff$.

%%%%%%%%%%%%%%%%%%%%%%%%%%%%%%%%%%%%%%%%%%%%%%%%%%%%%%%%%%%%%%%%%%%%%%%%%%%%%%%%%%%%%%%%%
%%%%%%%%%%%%%%%%%%%%%%%%%%%%%%%%%%%%%%%%%%%%%%%%%%%%%%%%%%%%%%%%%%%%%%%%%%%%%%%%%%%%%%%%%
%%%%%%%%%%%%%%%%%    Methods 
%%%%%%%%%%%%%%%%%%%%%%%%%%%%%%%%%%%%%%%%%%%%%%%%%%%%%%%%%%%%%%%%%%%%%%%%%%%%%%%%%%%%%%%%%
%%%%%%%%%%%%%%%%%%%%%%%%%%%%%%%%%%%%%%%%%%%%%%%%%%%%%%%%%%%%%%%%%%%%%%%%%%%%%%%%%%%%%%%%%

\section{Analysis and Results}\label{sec:methods_results}

%\subsection{Bayesian Analysis Method}

We present here our results for the odds ratios of the different prior distributions for $\chieff$ outlined in the previous section, as well as the posterior probability density functions for a ``mixed'' scenario with PBH providing a fraction $f$ of the BBH events. We then discuss how, under different assumptions, such odds ratios will evolve with additional events in the future, and how knowledge of which fraction of the events originates from which prior distribution will change with greater statistics (sec.~\ref{sec:future_events}).

%%%%%%%%%%%%%%%%%%%%%%%%%%%%%%%%%%%%%%%%%%%%%%%%%%%%%%%%%%%%%%%%%%%%%%%%%%%%%%%%%%%%%%%%%
%%%%%%%%%%%%%%%%%    Comparing models to observations 
%%%%%%%%%%%%%%%%%%%%%%%%%%%%%%%%%%%%%%%%%%%%%%%%%%%%%%%%%%%%%%%%%%%%%%%%%%%%%%%%%%%%%%%%%
\subsection{Comparing models to observations: odds ratios and mixture}

\begin{figure}
    \centering
    \includegraphics[width= 0.7\textwidth]{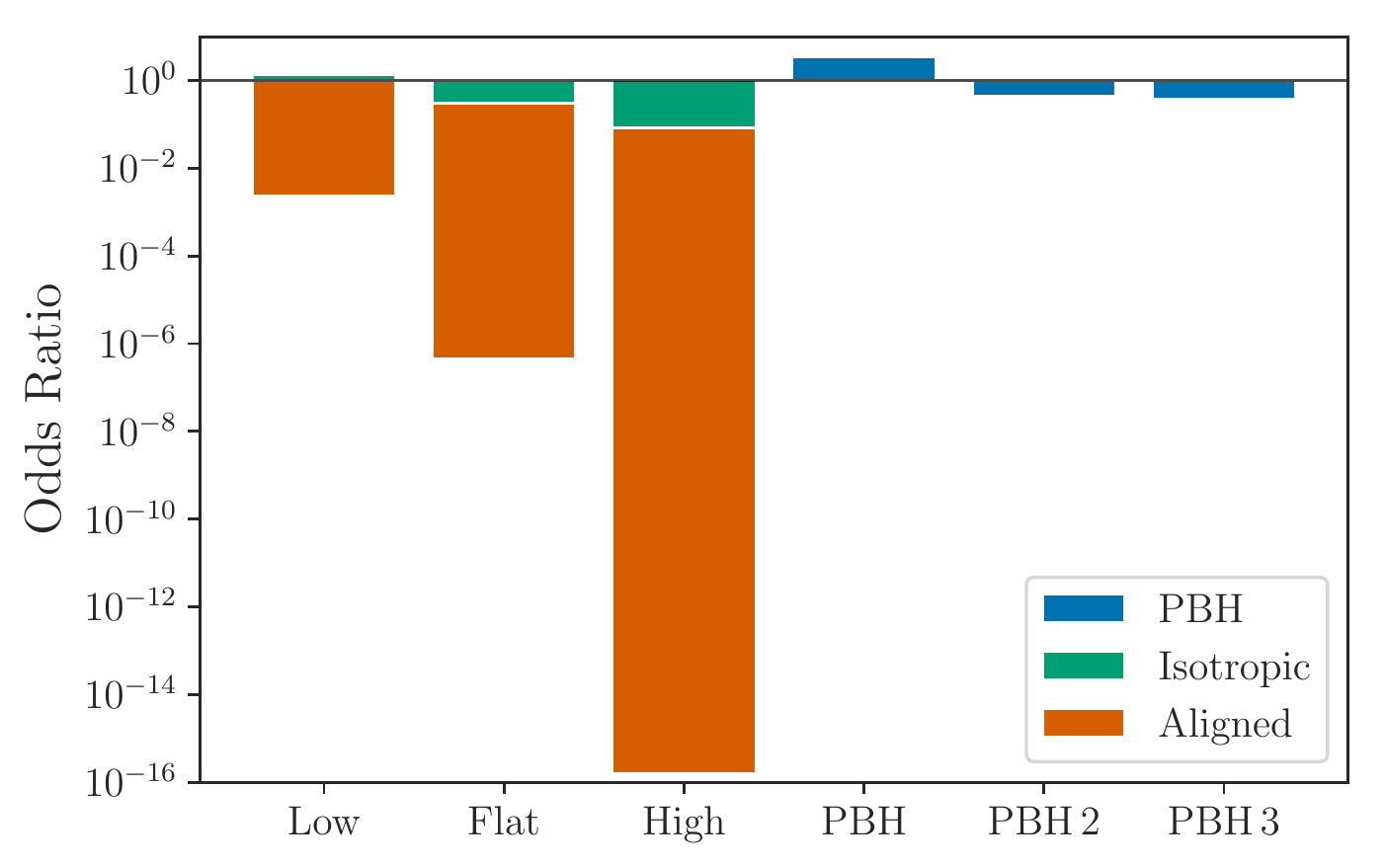}
    \caption{Odds ratios for different models with respect to the LI benchmark model. Larger odds ratios show higher statistical preference, with ratios larger than 1 indicating a preference with respect to the benchmark LI model. The low, flat and high spin magnitude are combined with the isotropic and aligned spin-orbit orientation distributions. The PBH model is for a fixed $\sigma=0.27$, PBH 2 and PBH 3 are for $(\nu=6,\gamma=0.8)$ and $(\nu=6$, $\gamma=0.88)$ respectively, in the notation of ref.~\cite{DeLuca:2019buf} (see text for details).}
    \label{fig:odds_10_events}
\end{figure}

\begin{table}
  \centering
\begin{tabular}{ l c c c }
LIGO &  \multicolumn{1}{c}{Low}   &  \multicolumn{1}{c}{Flat}   &  \multicolumn{1}{c}{High} \\  [0.2ex]  \hline   \hline  \\ [-1.4ex]
Isotropic  &  $0.0$   &  $-0.93$  & $-2.07$ \\ [0.14cm]
Aligned  &  $-4.12$  &  $-12.92$  & $-32.37$ \\ [0.14cm] \hline
\end{tabular}
\quad
\begin{tabular}{l c c c c}
This work & \multicolumn{1}{c}{Low}   &  \multicolumn{1}{c}{Flat} &  \multicolumn{1}{c}{High} & \multicolumn{1}{c}{PBH} \\  [0.2ex]  \hline   \hline  \\ [-1.4ex]
Isotropic  &  $0.0$   &  $-1.18$  & $-2.49$ & $0.39$ \\ [0.14cm]
Aligned   &  $-6.07$  &  $-14.65$  & $-36.41$ & \\ [0.14cm] \hline
\end{tabular}
\caption {Natural log Bayes factors for various spin distributions with $q=1$. \textit{Right}: Values reported by LIGO  \cite{LIGOScientific:2018jsj}. \textit{Left}: Values found in this work.}
\label{table: log_Bayes}
\end{table}

We confront here the prior distributions obtained in the previous section with data by calculating {\em odds ratios}, which quantify the statistical support for a model over another, allowing us to compare models and giving us a statistically-motivated selection criterion.
Fig.~\ref{fig:odds_10_events} shows the odds ratios between all possible models and the reference low-intrinsic-spin, isotropic (LI) model; what we show is therefore defined as $p(d|\mathrm{M})/p(d|\mathrm{LI})$ for model $\mathrm{M}$ given the 10 events $d$. Here, we have already assigned equal probability to the prior probability distributions of each and every model (i.e., we assume that all models are equally likely a priori). This implies that the odds ratio and the Bayes factor are equivalent. % fix the value of $\sigma = 0.27$ for the PBH model. 

Fig.~\ref{fig:odds_10_events} shows how all models with aligned spin distribution are significantly disfavored with respect to the isotropic ones. Also if we compare the ``favored'' aligned model, which is the one corresponding to a low intrinsic spin distribution (LA),  with the least favored isotropic model, which is the one with high intrinsic spin distribution (HI), the odds ratio in favour of HI is still very large, at $36:1$.

In addition to showing a strong statistical evidence for isotropy of spin and orbit orientations, the data favour models with small intrinsic spin magnitude distributions and heavily disfavour those with high spin. 
The two best models are the PBH and LI with the PBH model slightly preferred over the LI with an odds ratio of $3:2$.

Notice that the intrinsic spin distribution we assume for PBH was calculated in ref.~\cite{Chiba:2017rvs} by integrating the probability density $P(\chi,\delta)$, with $\chi$ the intrinsic spin and $\delta$ the overdensity giving rise to the PBH formation, over $\delta$. Critically, ref.~\cite{Chiba:2017rvs} assumes no correlation between $\delta$ and $\chi$. When this assumption is relaxed, one generally finds much narrower intrinsic spin distributions \cite{DeLuca:2019buf}. In the notation of ref.~\cite{DeLuca:2019buf}, the probability distribution for $\chi$ in the case of PBH depends on the parameter $\nu$, defined as the ratio of the critical collapse overdensity and the variance of the overdensity at horizon crossing, and on the parameter $\gamma$ which effectively measures the width of the PBH mass function, with $\gamma=1$ for a monochromatic power spectrum.

Figure 5 of ref.~\cite{DeLuca:2019buf} shows that the relevant range for the parameter $\nu$ for LIGO-sized PBH is around $\nu\sim6$. While the range for the parameter $\gamma$ depends on the PBH mass function, mass functions peaked around a few solar masses typically have values of $\gamma\sim0.85...0.88$ (see their sections 7.2 and 7.3), although a broader range is possible. Here, we take as benchmark cases $(\nu=6,\ \gamma=0.8)$ (which we indicate in Fig.~\ref{fig:odds_10_events} as PBH 2) and $(\nu=6,\ \gamma=0.88)$ (PBH 3 in Fig.~\ref{fig:odds_10_events}), the latter leading to the narrowest possible prior distribution for $\chieff$, and the former with a broader distribution. We show the prior distribution for the two models in the bottom panel of Fig.~\ref{fig:chieff_Priors_sigma_q}.

We find odds ratio of $0.44$ and $0.37$ respectively for PBH 2 and 3, indicating (since both odds ratios are smaller than 1) a statistical preference for the LI model as well as for our benchmark PBH model over these narrower intrinsic PBH spin distribution. As expected, our current ability to distinguish between models with very low spin distribution is very limited with the available data. Similar results and conclusions have been found in \cite{Farr:2017uvj,Tiwari:2018qch}. For reference and to summarize our findings, we list in Table \ref{table: log_Bayes} the natural log Bayes factors for various models compared to the benchmark LI model. 

\begin{figure}
\center\includegraphics[width=0.7
\textwidth]{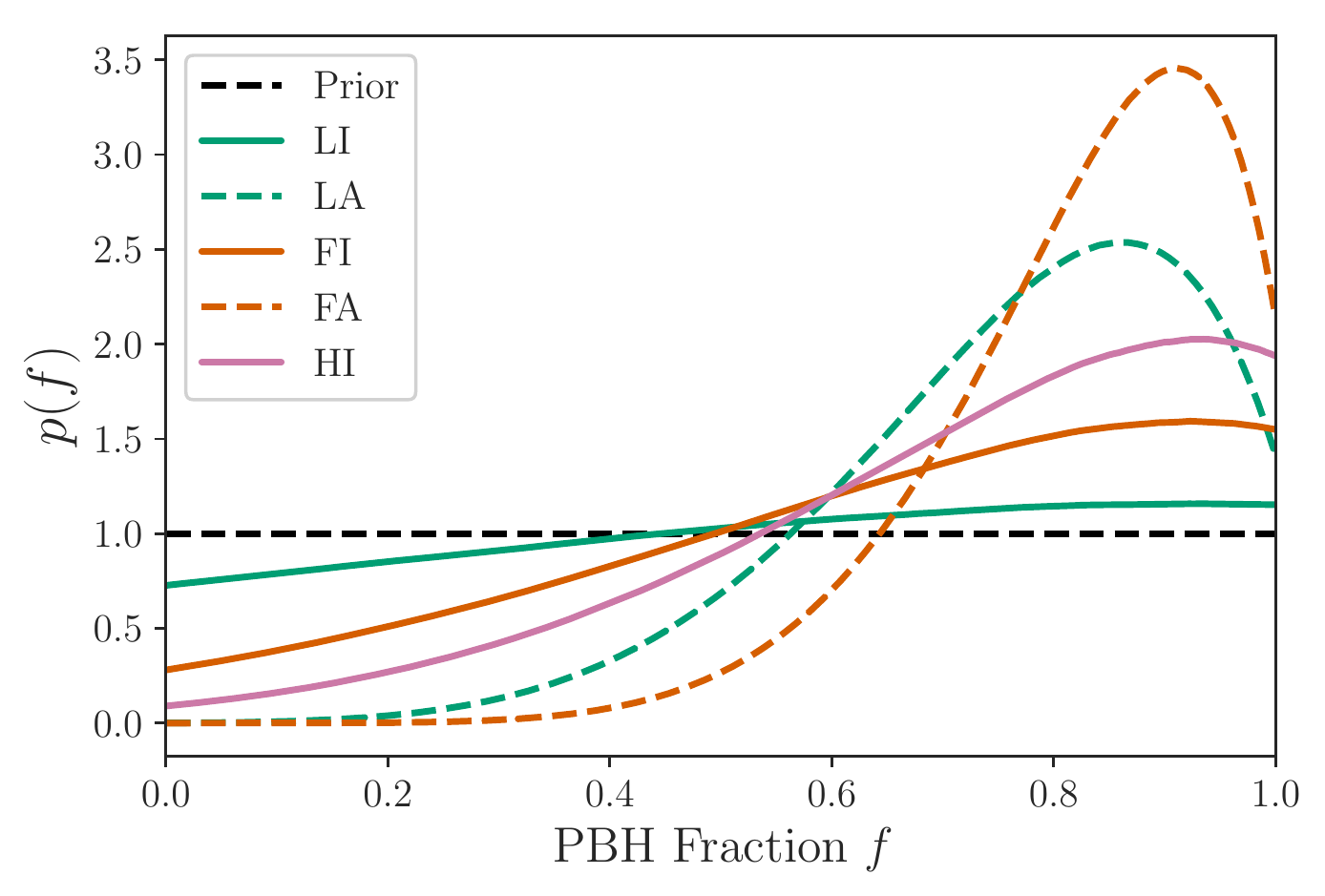}
\caption{Posterior probability density functions on the parameter $f$ for the 10 events observed by LIGO. $f=1$ corresponds to all coming from PBH.}
\label{fig:mix_10_ligo}
\end{figure}

The actual LIGO BBH population likely reflects a mixed population of two different models (or more). Here, we are going to assume that the mixture is our PBH model with the second one  any of the following models: LI, LA, FI, FA and HI. In Fig.~\ref{fig:mix_10_ligo} we show the posterior probability density for the {\em fraction of the BBH mergers coming from PBH}, which we indicate with $f$, where $f=1$ means that all the events are from PBH and $f=0$ means the opposite, i.e. all the events are coming from the second model and none from PBH. Notice that if we allow a mixed model we find a statistical preference for the majority of events coming from a PBH population. For the case where the mixed model consists of PBH and FA (dashed orange line), the $f$ distribution peaks around $\sim 0.9$, therefore favoring a scenario where 9 events come from a PBH population and 1 event from the FA population. This is somewhat expected because there clearly is one event (GW170729) that could have come more likely from a population favoring large $\chieff$ values such as what predicted in the FA prior distribution rather than PBH. A similar conclusion can be drawn when the second mixture model is LA (dashed green line). In this second case, there are {\em two events} that could be  ascribed to a LA distributions: GW170729 and GW170729. This is why the $f$ posterior distribution peaks around $\sim 0.85$, a little lower than the case of FA. For the case of a mixture of FI (solid orange line) or HI (solid pink line) with PBH, the probability distribution for $f$ is flatter, but we still can conclude that more than half of the events are coming from a PBH population. Lastly, in the case of a mixture model consisting of PBH and LI (solid green line) the distribution is almost half PBH and half LI with a slight preference for the PBH model, as expected from the odds ratio between the two models, which have comparably similar $\chieff$ prior distributions.

%%%%%%%%%%%%%%%%%%%%%%%%%%%%%%%%%%%%%%%%%%%%%%%%%%%%%%%%%%%%%%%%%%%%%%%%%%%%%%%%%%%%%%%%%
%%%%%%%%%%%%%%%%%    Future Events
%%%%%%%%%%%%%%%%%%%%%%%%%%%%%%%%%%%%%%%%%%%%%%%%%%%%%%%%%%%%%%%%%%%%%%%%%%%%%%%%%%%%%%%%%

\subsection{Future events}\label{sec:future_events}
To test the sensitivity of our setup to the different benchmark models under consideration, we simulate future events for each of the six population models under consideration: PBH, LI, LA, FI, FA and HI. We generate mock observations following the same approach as in refs.~\cite{Farr:2017gtv, Farr:2017uvj}. First, We approximate each LIGO event as a Gaussian with the same mean value and $90\%$ credible interval; Second, we draw a value of $\chieff^{\mathrm{true}}$ from the population's distribution we want to simulate; third, we  generate an observation from the distribution $\chieff^{\mathrm{obs}} \sim \mathcal{N}(\chieff^{\mathrm{true}}, \sigma^{\mathrm{unc}})$, where $\sigma^{\mathrm{unc}}$ is a random uncertainty from one of LIGO's ten events. Finally, the we calculate the posterior probability as $\sim \mathcal{N}(\chieff^{\mathrm{true}}, \sigma^{\mathrm{unc}})p_{\mathrm{FI}}(\chieff)$. 
\begin{figure}
    \centering
    \includegraphics[width= .48\textwidth]{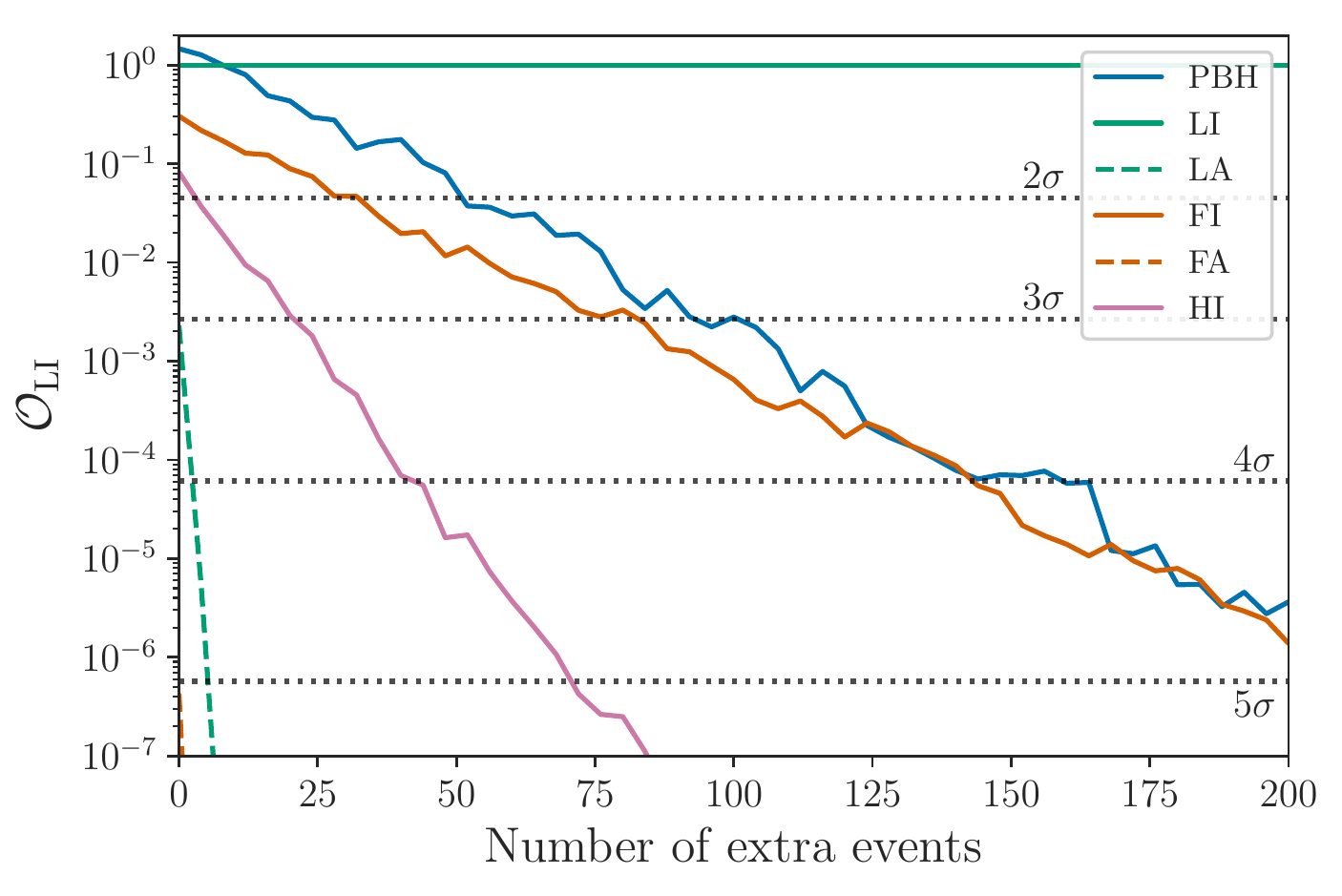}
    \includegraphics[width= .48\textwidth]{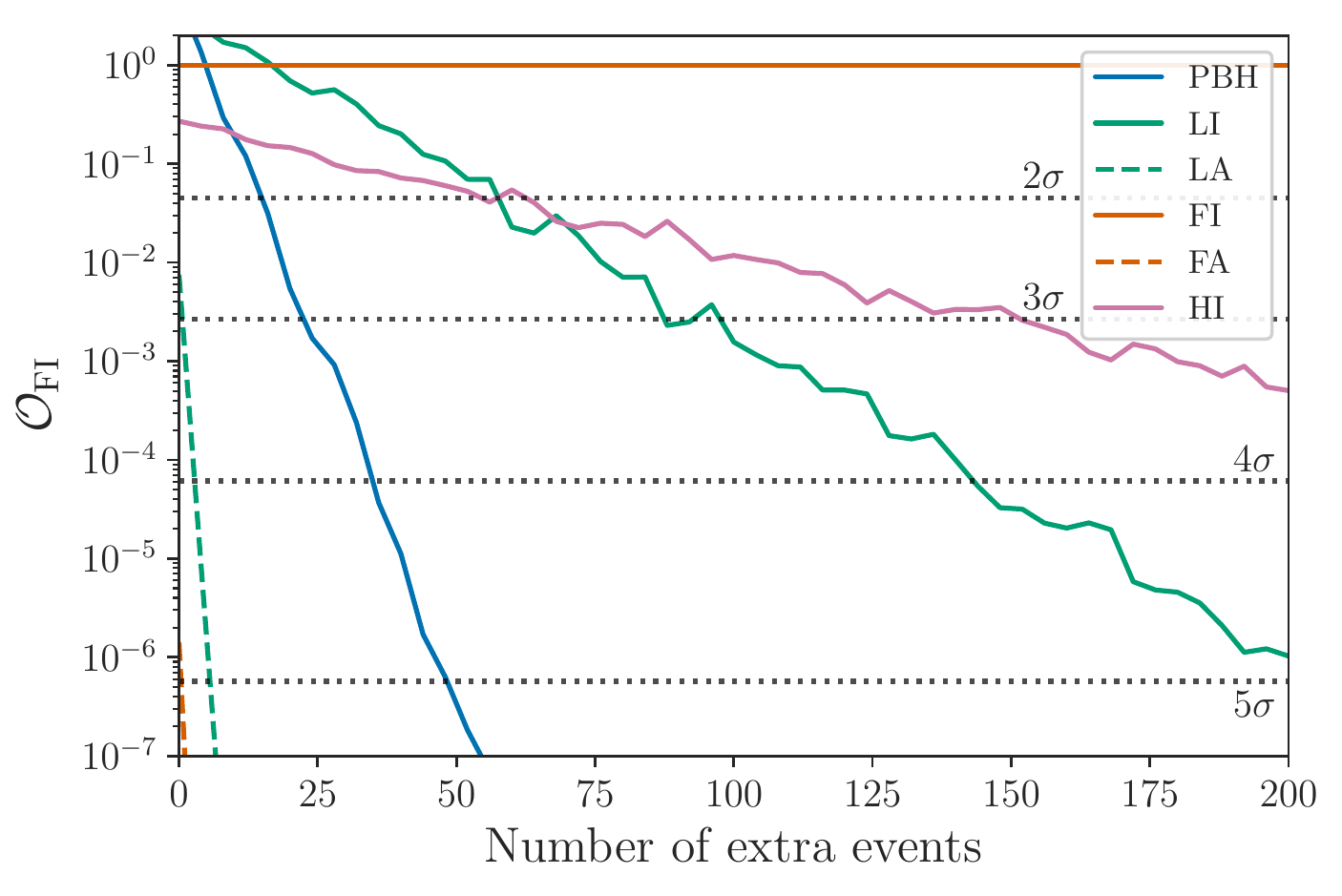}
    \caption{Evolution of the odds ratio as a function of the number of extra events for 200 LI (\textit{left}) and 200 FI (\textit{right}) simulated events.}
    \label{fig:LI_FI_odds}
\end{figure}

\begin{figure}
    \centering
    \includegraphics[width= 0.7\textwidth]{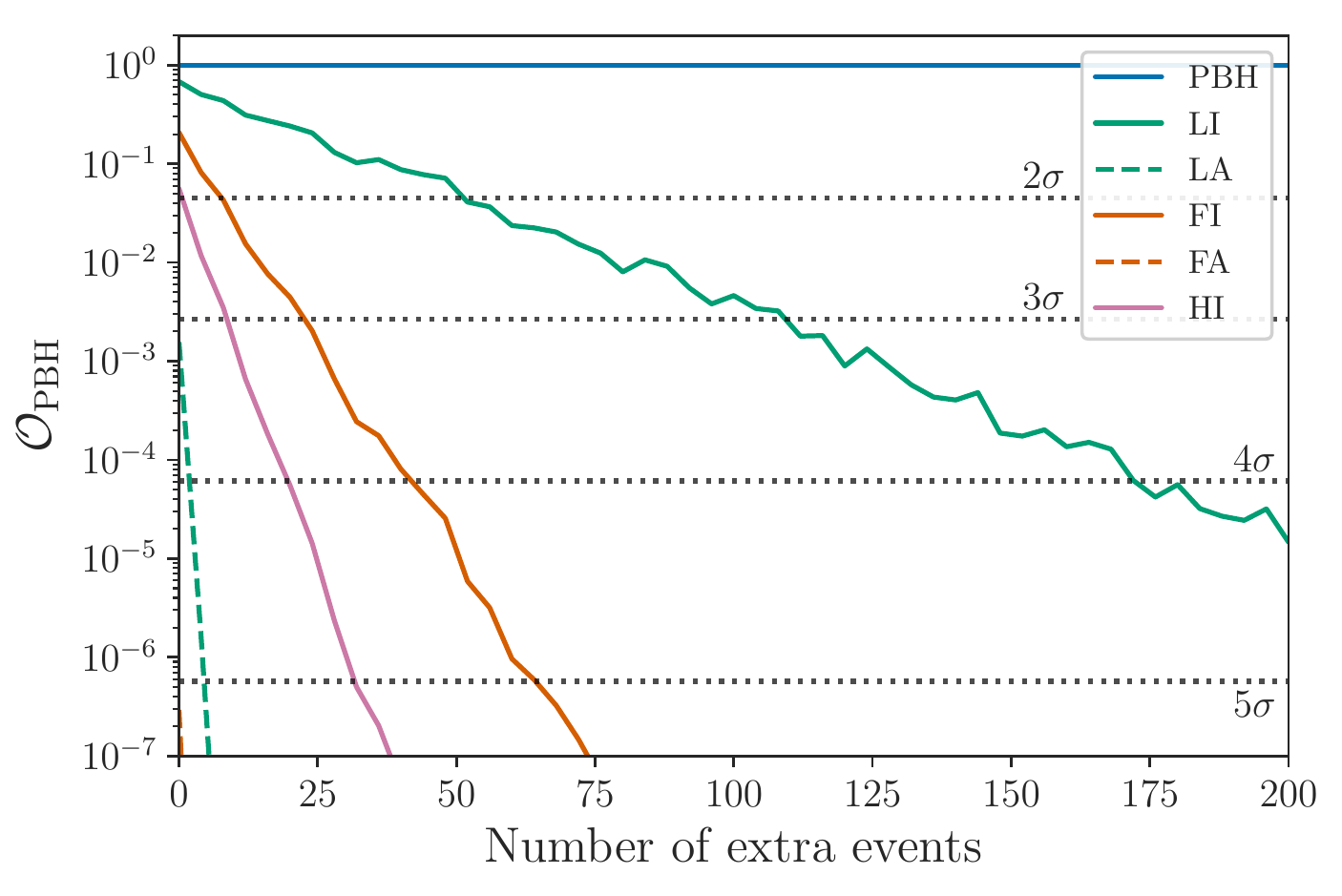}
    \caption{Evolution of the odds ratio as a function of the number of extra events for 200 PBH simulated events.}
    \label{fig:PBH_odds}
\end{figure}

Assuming that all events are coming from the same population, we simulate 200 events for three possible ``true'' scenarios: FI, LI or PBH.  Fig.~\ref{fig:LI_FI_odds} shows the dependence of the odds ratio with respect to the number of extra events from LI (left) and FI (right) populations. Notice that the odds ratios for the models already start at different values, because the current 10 LIGO events are included: the starting point for each model is thus just the odds ratio from Table \ref{table: log_Bayes}. 

For the fully LI-simulated population, our results show that with only 10 extra events the FA and LA models might be disfavoured at more than the $5\sigma$ level, and that 75 extra events are needed to reject the the HI model at the same confidence level.  The LI and PBH models would be disfavored at the same level with even more events, reaching close to a $5 \sigma$ level with 200 extra events.

Under the assumption that the true population is FI, the evolution of the odds ratios is shown in the right panel of Fig.~\ref{fig:LI_FI_odds}. With less than 10 extra events we find that it would be possible to heavily disfavor both the FA and LA models; interestingly, the PBH model can also be rejected at the $5\sigma$ level with only 50 extra events. The entire 200 extra events would allow for a $5 \sigma$ rejection threshold for the LI model and $3\sigma$ for the HI model. Finally, if the true population is that of PBHs, as in Fig.~\ref{fig:PBH_odds}, all models except LI can be rejected  at $5 \sigma$ level with only 75 extra events. We find that 200 extra events would be needed to discriminate the LI model over the PBH one at the $4\sigma$ level.

\begin{figure}
    \centering
    \includegraphics[width= .48\textwidth]{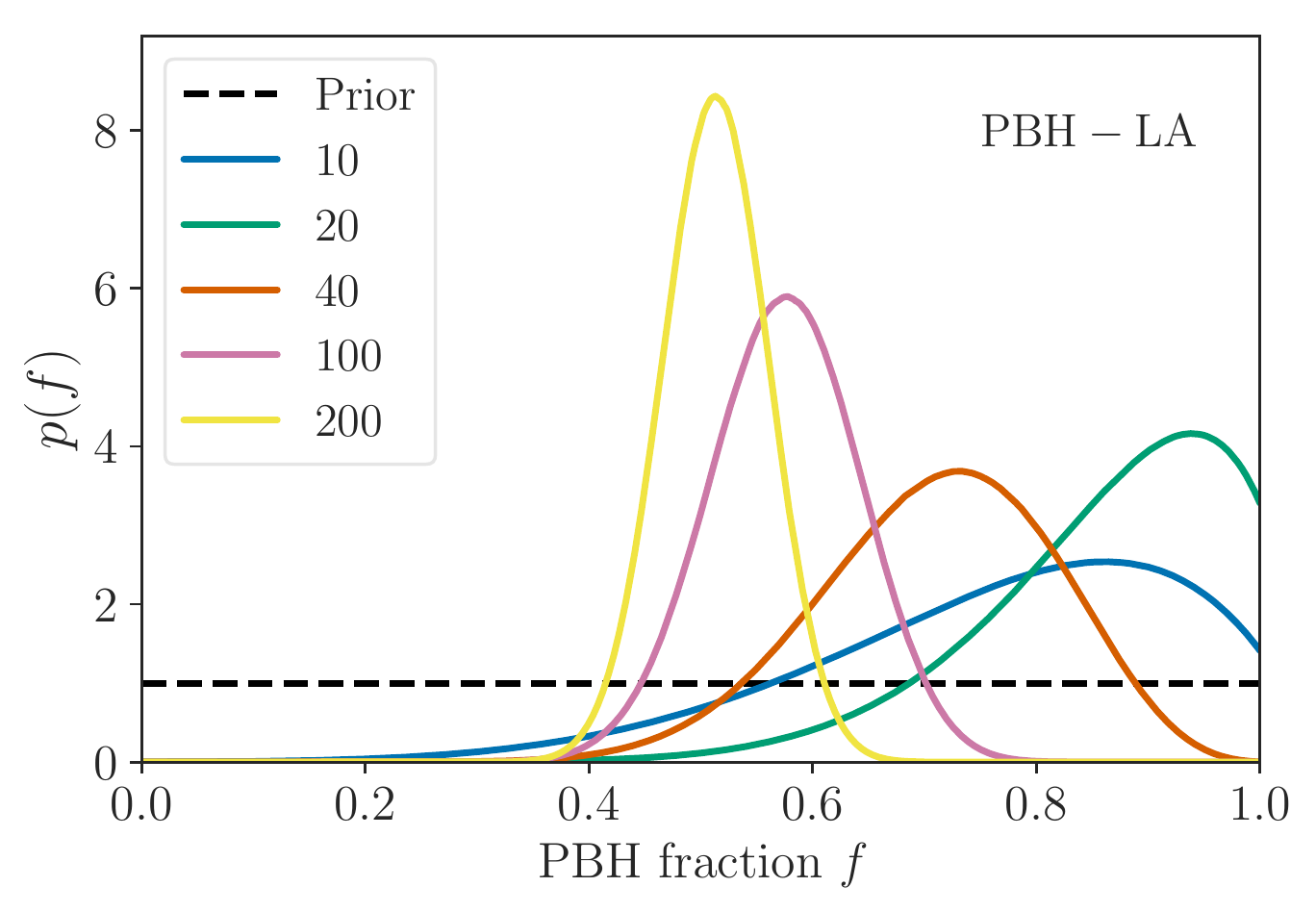}
    \includegraphics[width= .48\textwidth]{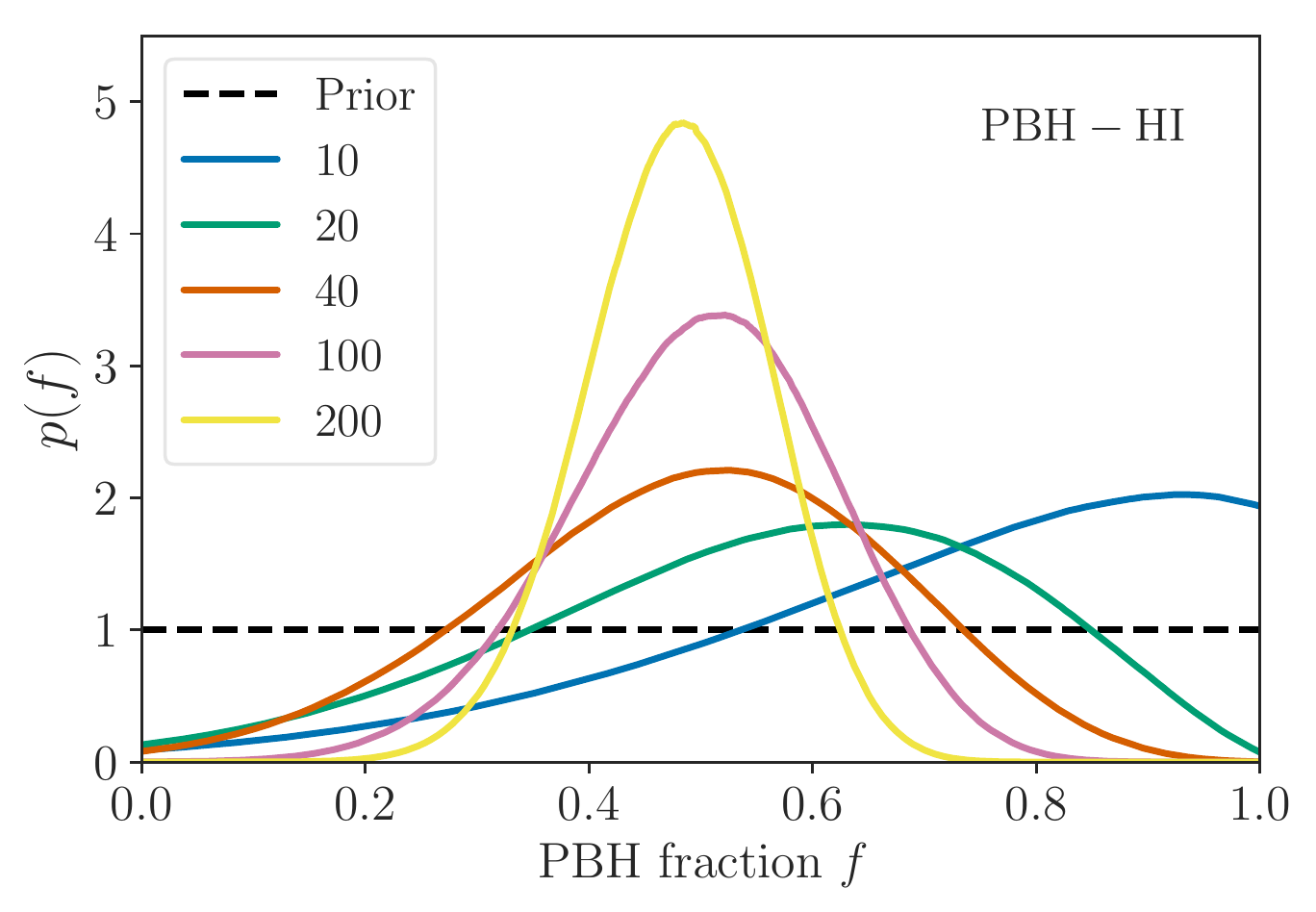}
    \includegraphics[width= .48\textwidth]{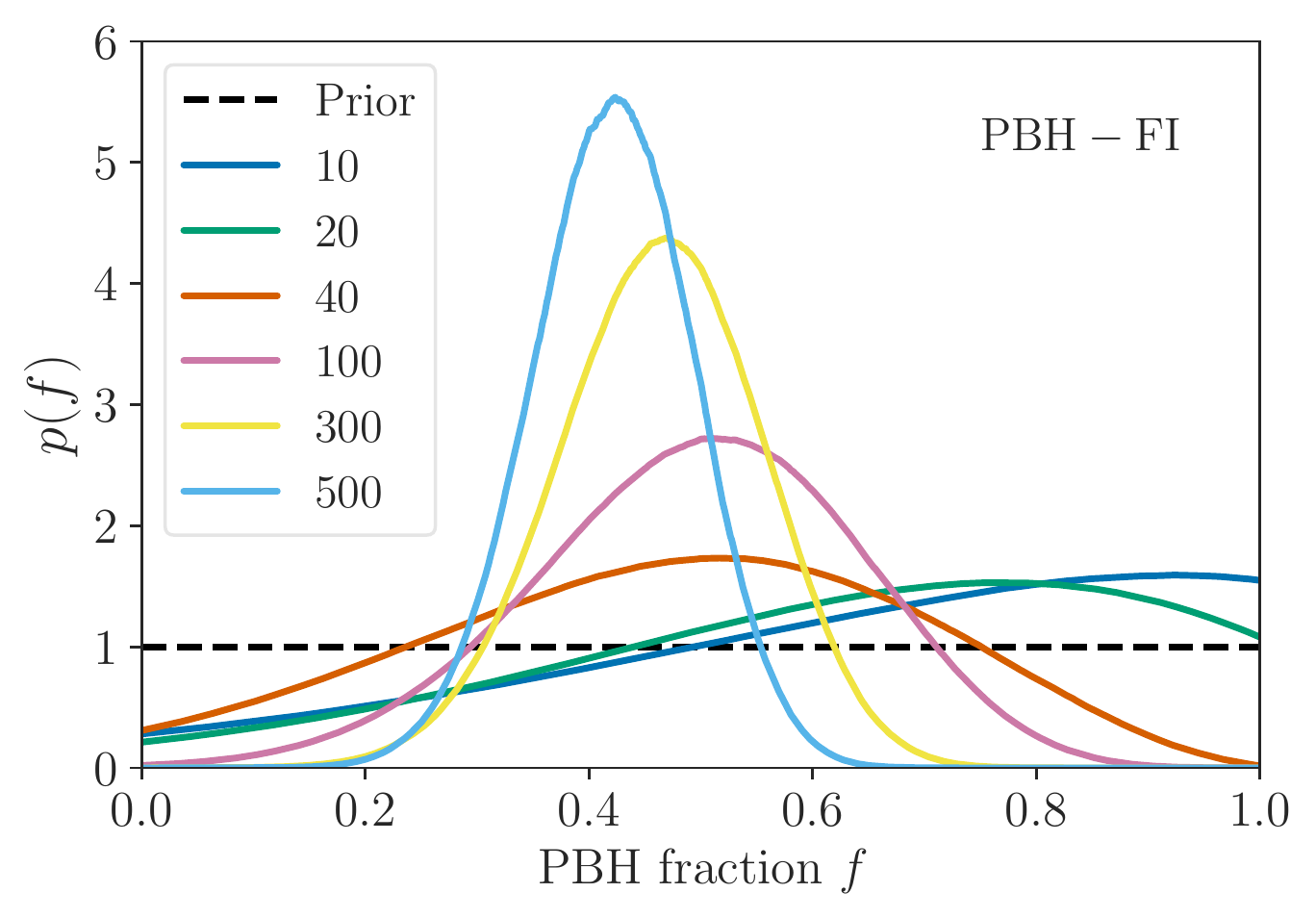}
    \includegraphics[width= .48\textwidth]{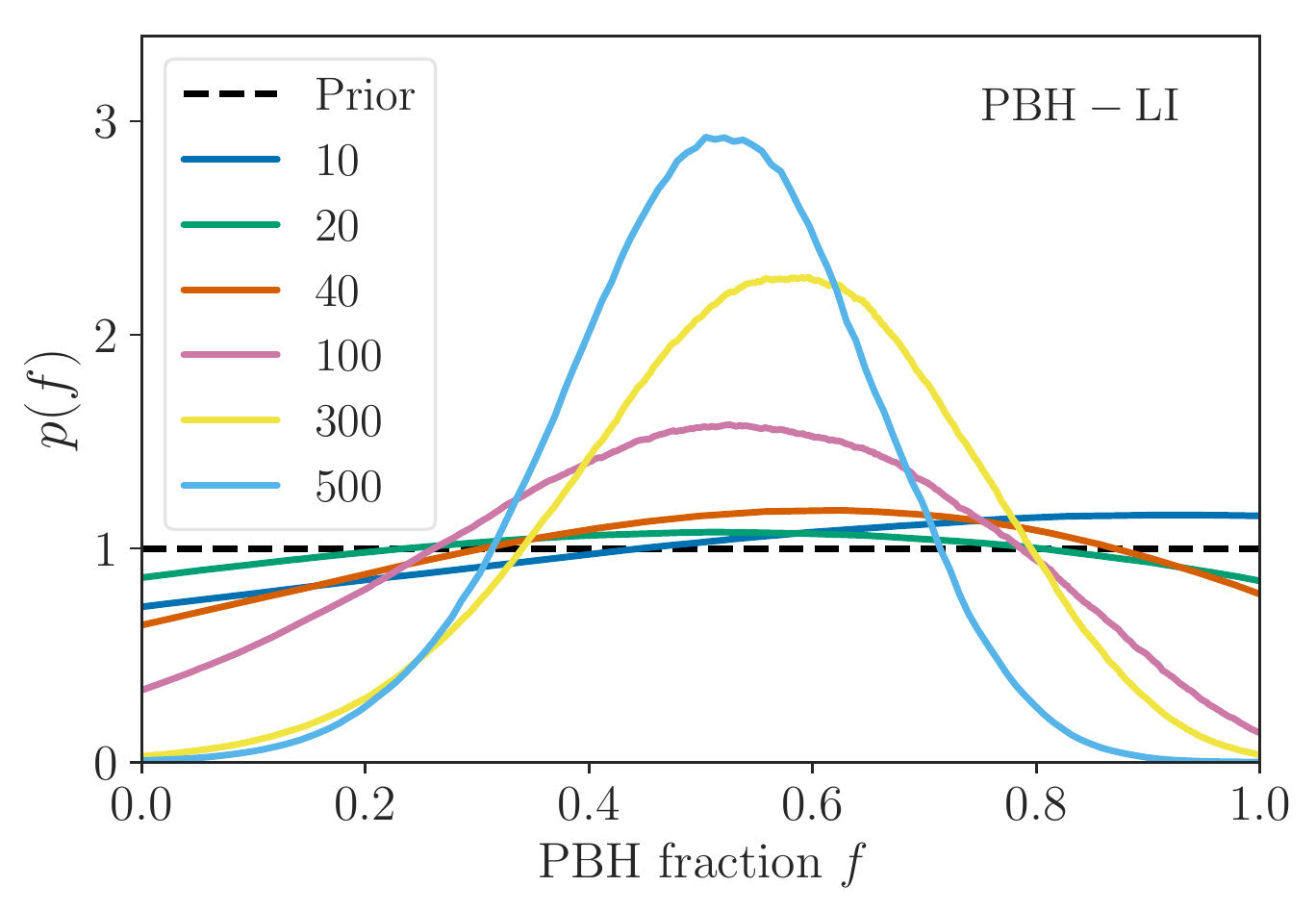}
    \caption{Posterior probability density functions for the PBH fraction $f$ for 4 
 mixed populations.}
    \label{fig:Mix_models}
\end{figure}

Once again, we emphasize that there is no reason to assume all events observed by LIGO are coming from the same single population. In what follows we therefore consider a mixed population consisting of half PBH and half of a second model among LI, FI, HI or LA. We simulate up to 500 mixed events for each of these mixed populations. 

In the upper panels of Fig.~\ref{fig:Mix_models}, we show that it is possible to infer the relative fraction of PBH $f$ (which is given on the horizontal axis) with high confidence with 200 events for the case of PBH-LA and PBH-HI mixture models. To be able to discriminate the mixture fraction between PBH and FI, more events are needed than in the previous two cases, as shown in the bottom left panel of Fig.~\ref{fig:Mix_models}. After detecting 500 events the value of $f$ can be determined fairly precisely. Additional events are needed when the two prior distributions for the two models are similar, such as for the case of a mixed PBH-LI population, bottom right in Fig.~\ref{fig:Mix_models}. This notwithstanding, the value of $f$ peaks at the correct value of $\sim 0.5$ but the spread is still substantial,  even with 500 extra events. 

Run O3 of LIGO-Virgo has already commenced, and the projected inferred rate of BBH mergers from the previous runs is around $9.7-101 \, \rm{Gpc}^{-3}\, \rm{y}^{-1}$ \cite{LIGOScientific:2018mvr}. At this moment, the number of putative candidate run O3 BBH merger events is 10, from a period of approximately 1.5 months\footnote{\href{https://gracedb.ligo.org/latest/}{https://gracedb.ligo.org/latest/}}, which would imply an approximate total number of events per year of around 80.

%%%%%%%%%%%%%%%%%%%%%%%%%%%%%%%%%%%%%%%%%%%%%%%%%%%%%%%%%%%%%%%%%%%%%%%%%%%%%%%%%%%%%%%%%
%%%%%%%%%%%%%%%%%    Discussion and Conclusions 
%%%%%%%%%%%%%%%%%%%%%%%%%%%%%%%%%%%%%%%%%%%%%%%%%%%%%%%%%%%%%%%%%%%%%%%%%%%%%%%%%%%%%%%%%
\section{Discussion and Conclusions}\label{sec:conclusions}

In this study we considered how measurements of the effective spin parameter $\chieff$ provide information on the origin of merging binary black holes observed with gravitational wave telescopes. To this end, we presented a calculation of the prior distribution for primordial black holes as well as for a few representative benchmark distributions possibly indicative of what expected in simple astrophysical black hole binary populations.

In the case of PBHs, following ref.~\cite{Chiba:2017rvs} we assumed no correlation between the overdensity leading to the black hole formation in the early universe and the intrinsic black hole spin; the resulting intrinsic spin distribution is a positive-definite half-Gaussian with zero mean; to fix the width of the intrinsic spin distribution we calculated the prior distribution for the effective spin parameter $\chieff$ and we utilized 10 LIGO-Virgo measurements of $\chieff$ to determine a best-fit intrinsic spin width.

We then proceeded to compare odds ratios for the current set of 10 measurements for $\chieff$ for the various models under consideration. We showed that with current data the $\chieff$ measurements have a marginal, and not highly statistical significant preference for a dominant population of primordial black holes over the best fitting astrophysical model. We also calculated the posterior probability for the relative fraction of primordial versus astrophysical BBH events, finding that there is a preference for a scenario where one or two events originate from a population with preferentially aligned spin-orbit distributions, and the remaining eight-to-nine events from a population with an isotropic spin-orbit distribution and low intrinsic spin.

We then assessed the number of future events needed to disentangle, at a given significance level, BBH from different populations, assuming that all binaries originate from the same ``true'' distribution. Generally, non-isotropic spin distributions are highly disfavored even when only considering the current 10 events. In addition, even for isotropic alignment distributions, it will quickly become possible to distinguish models with large intrinsic spin from those with low intrinsic spin magnitude distributions. Assuming that the entirety of the merging black holes have a primordial origin, we anticipate that distinguishing at 3$\sigma$ their $\chieff$ distribution from a low intrinsic spin, aligned spin distribution will require on the order of 100 additional events.

Finally, we illustrated the number of events necessary to acquire information on the relative fraction of primordial versus non primordial binary black holes. Once again, if the population of non-primordial black holes has a preferentially aligned spin-orbit distribution, such fraction can be pinpointed with relatively few additional events; the posterior distribution for the relative fraction of events from populations with isotropic spin-orbit distribution shows a peak generally pointing to a systematically larger-than-true fraction of primordial black holes, but eventually converging with large-enough statistics to the ``true'' value.

In the Appendix, we discuss the systematics associated with three key assumptions used in our analysis: (1) the width $\sigma$ of the primordial black hole intrinsic spin distribution, (2) the mass ratio we use to calculate the $\chieff$ prior distribution, and (3) the assumption of perfect alignment between spin and orbit.

Clearly, as an increasingly large statistics of BBH mergers becomes available, the question of the origin of the population of merging black holes will come into sharper focus not only with studies of the effective spin parameter, but using all other pieces of information, including but not limited to the mass distribution of events, the correlation between mass and spin, localization information, etc. Interesting questions relating for instance to the degree to which BBH accrete will also be tackled, including the issue of whether there could be substantial accretion for PBH and how much accretion would spin up individual PBHs \cite{Postnov:2019tmw}: possibly, if PBHs do significantly accrete, the heavier ones would have both higher spin and higher $\chieff$. Interestingly, this seems to be the case for the most massive LIGO event (GW170729) and for the recent, claimed detection events GW151216  \cite{Zackay:2019tzo} and GW170403 \cite{Venumadhav:2019lyq}. The central issue of observational bias in inferring the origin of merging BBH will also be helped greatly by benefiting from increased statistics. Finally, we also expect that theoretical and observational progress will lead to more realistic models for the expected distributions for the intrinsic spin and spin-orbit correlation of different populations of astrophysical black holes, allowing firmer statements on the origin of merging black holes than what is possible with the benchmark models currently in use.

%%%%%%%%%%%%%%%%%%%%%%%%%%%%%%%%%%%%%%%%%%%%%%%%%%%%%%%%%%%%%%%%%%%%%%%%%%%%%%%%%%%%%%%%%
%%%%%%%%%%%%%%%%%    Acknowledgments
%%%%%%%%%%%%%%%%%%%%%%%%%%%%%%%%%%%%%%%%%%%%%%%%%%%%%%%%%%%%%%%%%%%%%%%%%%%%%%%%%%%%%%%%%

\section*{Acknowledgments}
We are grateful to Yonatan Kahn and to Benjamin V. Lehmann, John Tamanas and other members of the UCSC particle theory group for helpful comments. This work is partly supported by the U.S.\ Department of Energy grant number de-sc0010107.
\appendix

%%%%%%%%%%%%%%%%%%%%%%%%%%%%%%%%%%%%%%%%%%%%%%%%%%%%%%%%%%%%%%%%%%%%%%%%%%%%%%%%%%%%%%%%
%Priors
%%%%%%%%%%%%%%%%%%%%%%%%%%%%%%%%%%%%%%%%%%%%%%%%%%%%%%%%%%%%%%%%%%%%%%%%%%%%%%%%%%%%%%%%
\section{Priors}

In this Appendix we derive the probability distribution for $\chieff$ for general given {\em spin magnitude} and {\em angular} distributions. Subsequently, we show how the distribution for $\chieff$ differs when we relax some assumptions that we employed in our calculations, specifically the assumption that the merging black holes have the same mass, and the assumption of perfect alignment between the individual spin and the orbital angular momentum for the ``aligned'' models. 

Let us define the direction of the orbital angular momentum as the direction of the $z${-}axis, then $\chieff$ can be re-written as 
\beq
\chieff = \frac{ \chi^{z}_1 + q \chi^{z}_2 }{1+q},
\eeq
where $q$ is the mass ratio $m_{2}/m_{1}$ such that $0 \leq q \leq 1$, and $\chi^{z}_i=\chi_{i} \cos \theta_{i}$ is the individual spin component along the $z${-}axis. Our goal is to derive the distribution for $\chi^{z}_i$ and the probability distribution for the sum $\chi^{z}_1 + q \chi^{z}_2$. 

First, the distribution of the product for $\chi^{z}_{i}$ is given by the following integral over probability distributions
\beq
p(\chi^{z}_{i})= \int_{0}^{1} f_{\chi_{i}}(\chi_{i}) \, \dd \chi_{i} \int_{-1}^{1} f_{\cos \theta_{i}}(\cos \theta_{i})\, \delta(\chi^{z}_{i} - \chi_{i}\,\cos \theta_{i} )\, \dd \cos \theta_{i} ,
\eeq
where the probability density functions $f_{\chi_{i}}$ is just the spin magnitude distribution, as given, for example, for PBH in Eq.~(\ref{Eq:PBH_Probability_chi}), and where $f_{\cos \theta_{i}}$ is the distribution for the cosine of $\theta_{i}$ ($\theta_{i}$ is defined under Eq.~(\ref{eq:chieff})). For instance, if the angular distribution is isotropic the $p(\chi^{z})$ distribution reads as
\beq
p(\chi^{z}_{i})= \frac{1}{2}\int_{\left| \chi^{z}_{i} \right|}^{1} f_{\chi_{i}}(\chi^{z}_{i}/\cos \theta_{i})\dfrac{1}{|\cos \theta_{i}|} \dd \cos \theta_{i} .
\eeq
Using the probability distribution for $\chi^{z}_{i}$, and assuming that the individual spins are independent of each other, we find that the probability distribution for $\chieff$ is given by the following convolution of the distributions
\beq
\label{Eq:Prior_Chieff}
p\left(\chieff \right) & = \int_{-1}^1 \dd \chi^{z}_1  \int_{-1}^1  \delta\left( \chieff - \frac{\left(\chi^{z}_1 + q \chi^{z}_2 \right)}{(1+q)} \right) p\left( \chi^{z}_1\right) p\left(\chi^{z}_2\right) \dd \chi^{z}_2 \\
& = (1+q)\int_{a}^{b}  \, p\left(\chi^{z}_2\right) p\left((1+q)\chieff - q \chi^{z}_2 \right)  \dd \chi^{z}_2 \,,
\eeq
where the extrema of integration read
\beq
a&=\max\left(-1, \frac{(1+q)\chieff - 1}{q} \right)\\
b&=\min\left(1, \frac{(1+q)\chieff + 1}{q}\right).
\eeq
In general, there is no analytic solution for the probability density, and we have to perform the integration numerically to solve for the distribution.

\begin{figure}
    \centering
    \includegraphics[width= .48\textwidth]{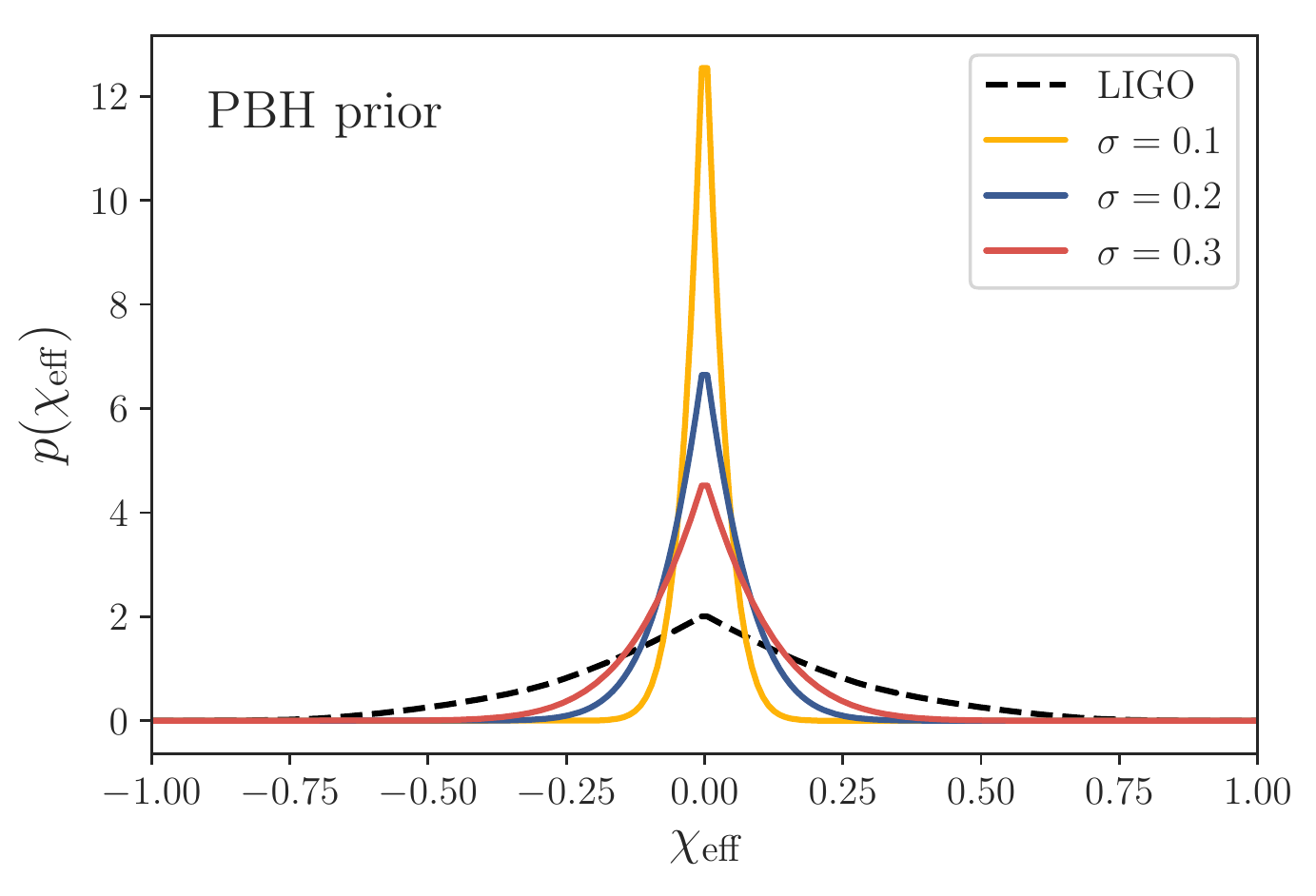}
    \includegraphics[width= .48\textwidth]{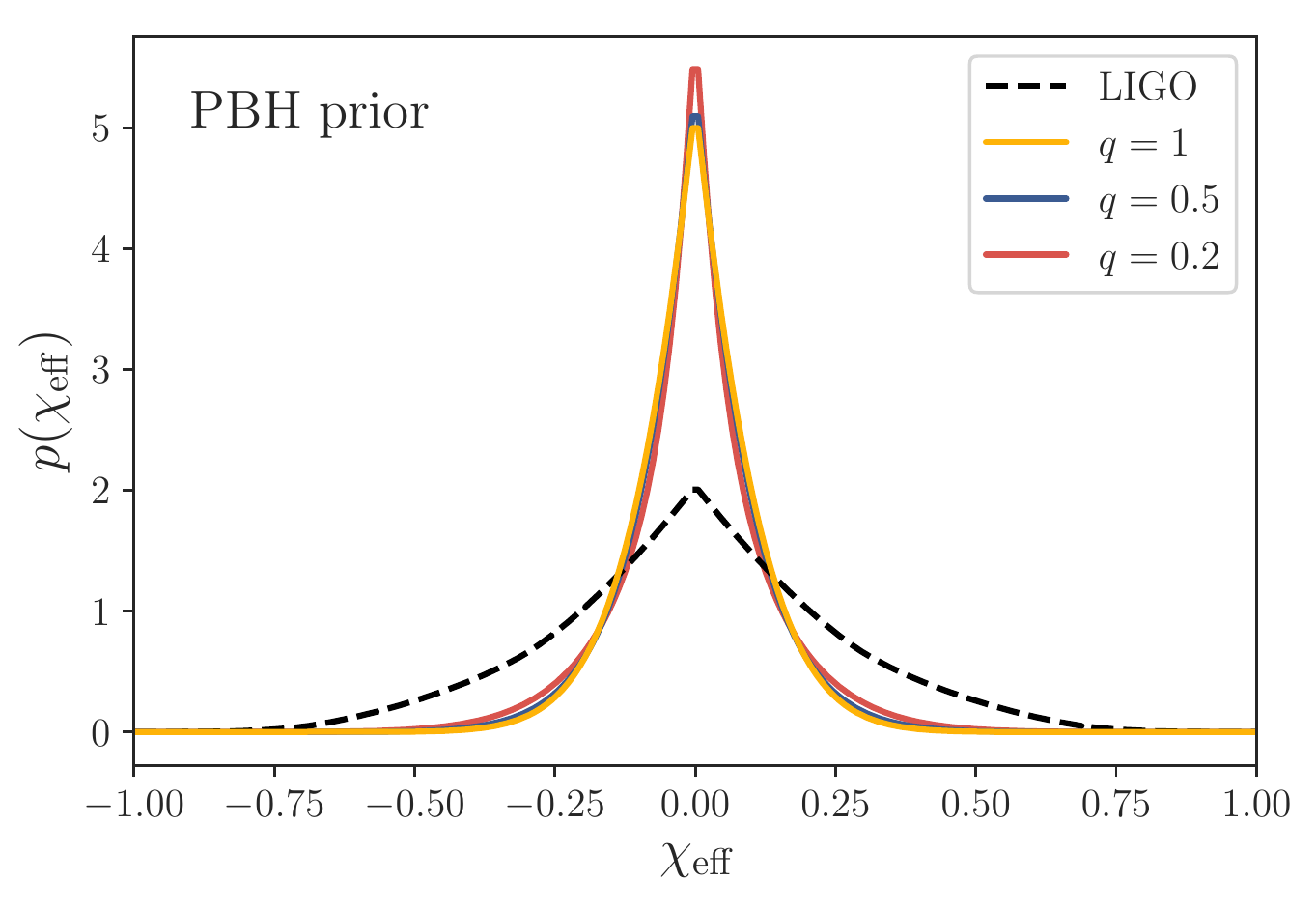}
    \includegraphics[width= .48\textwidth]{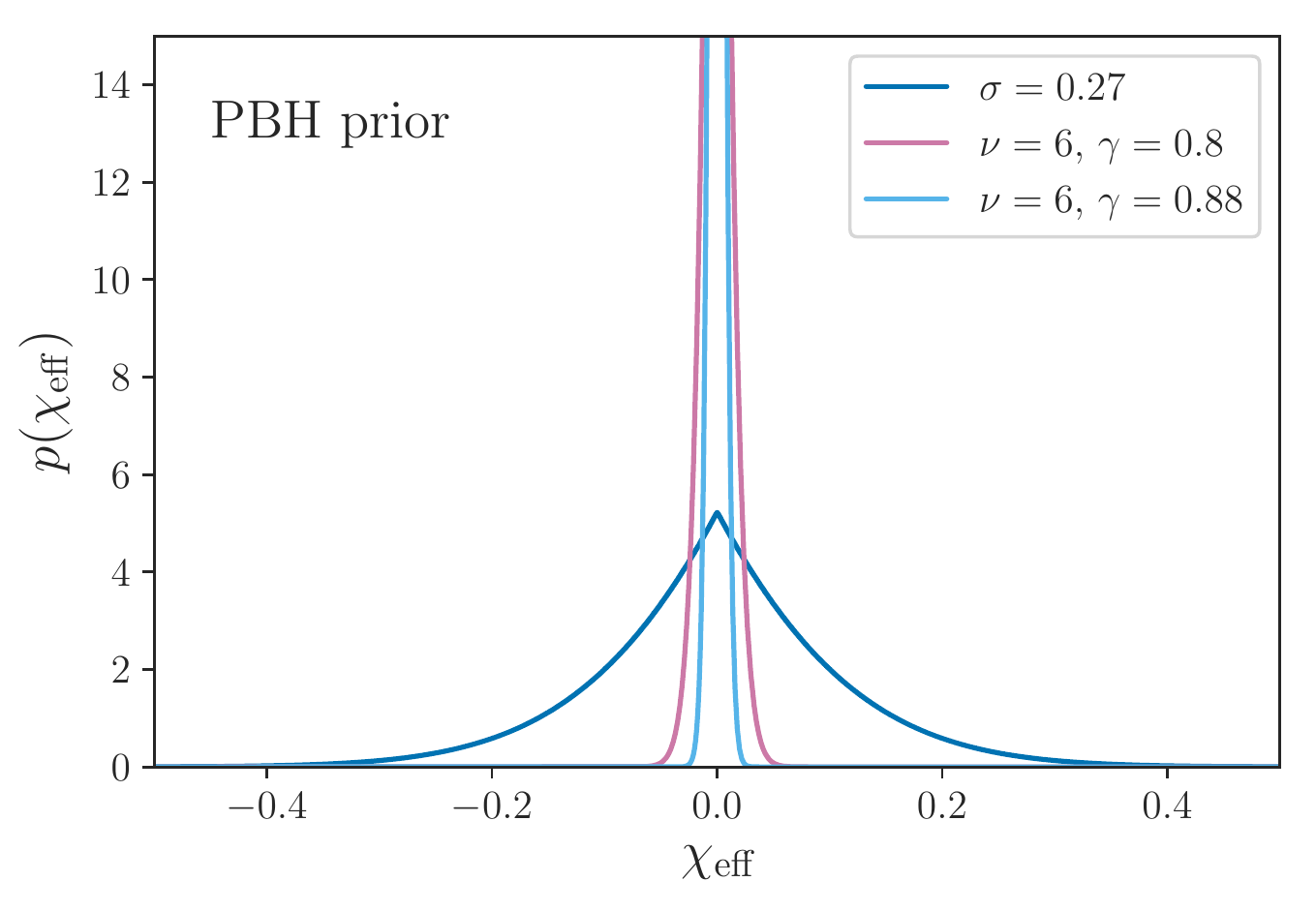}
    \caption{\textit{Top left}: Prior distribution for $\chieff$ used in LIGO analysis and for the PBH model with various values of $\sigma$. \textit{Top right}: Prior distribution for $\chieff$ for different mass ratios $q=1$, $q=0.5$ and $q=0.2$. \textit{Bottom}: Prior distribution for $\chieff$ for different PBH spin distribution (see sec. \ref{sec:PBH Spin Distribution} for details).}
    \label{fig:chieff_Priors_sigma_q}
\end{figure}

\begin{figure}
    \centering
    \includegraphics[width= .48\textwidth]{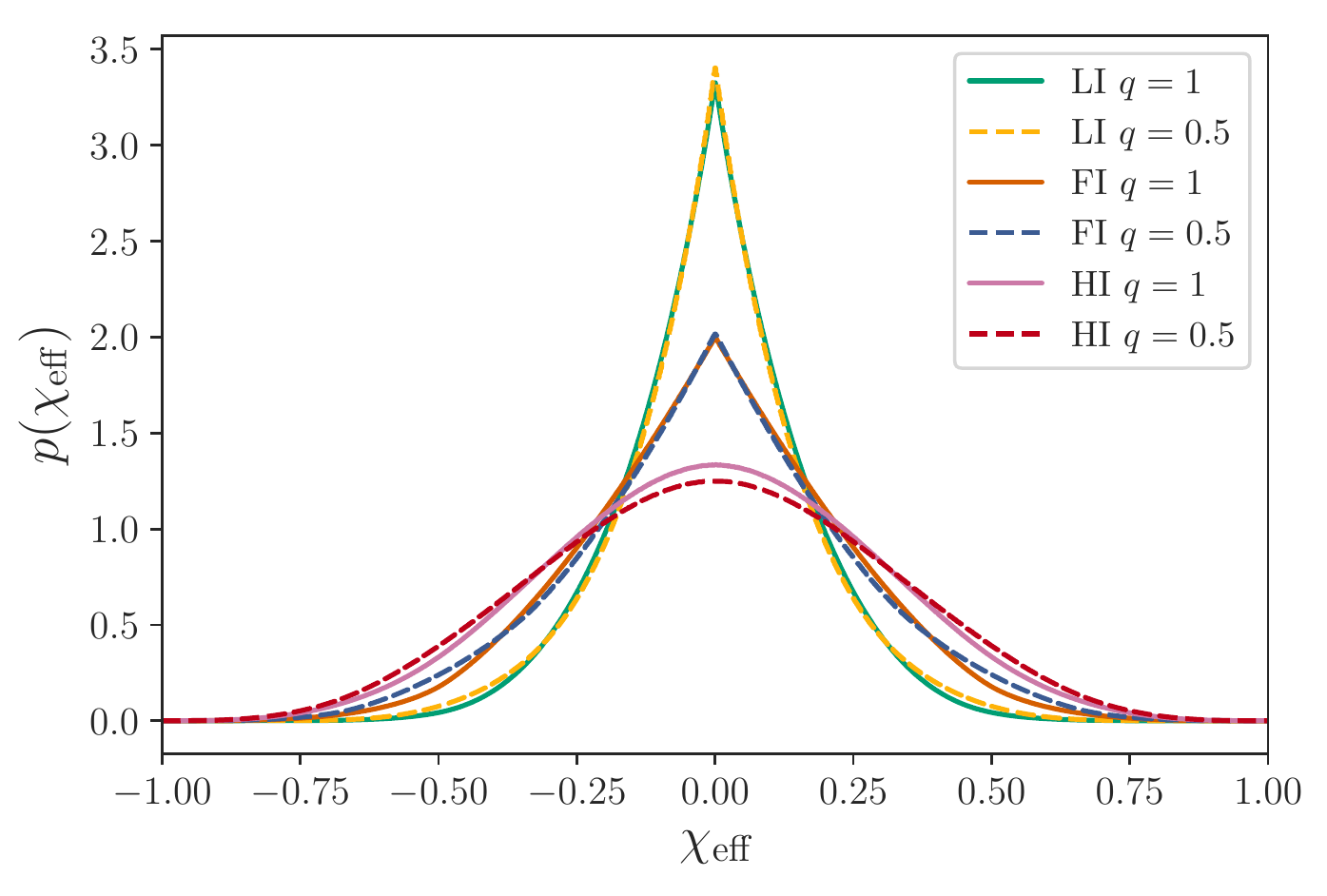}
    \includegraphics[width= .48\textwidth]{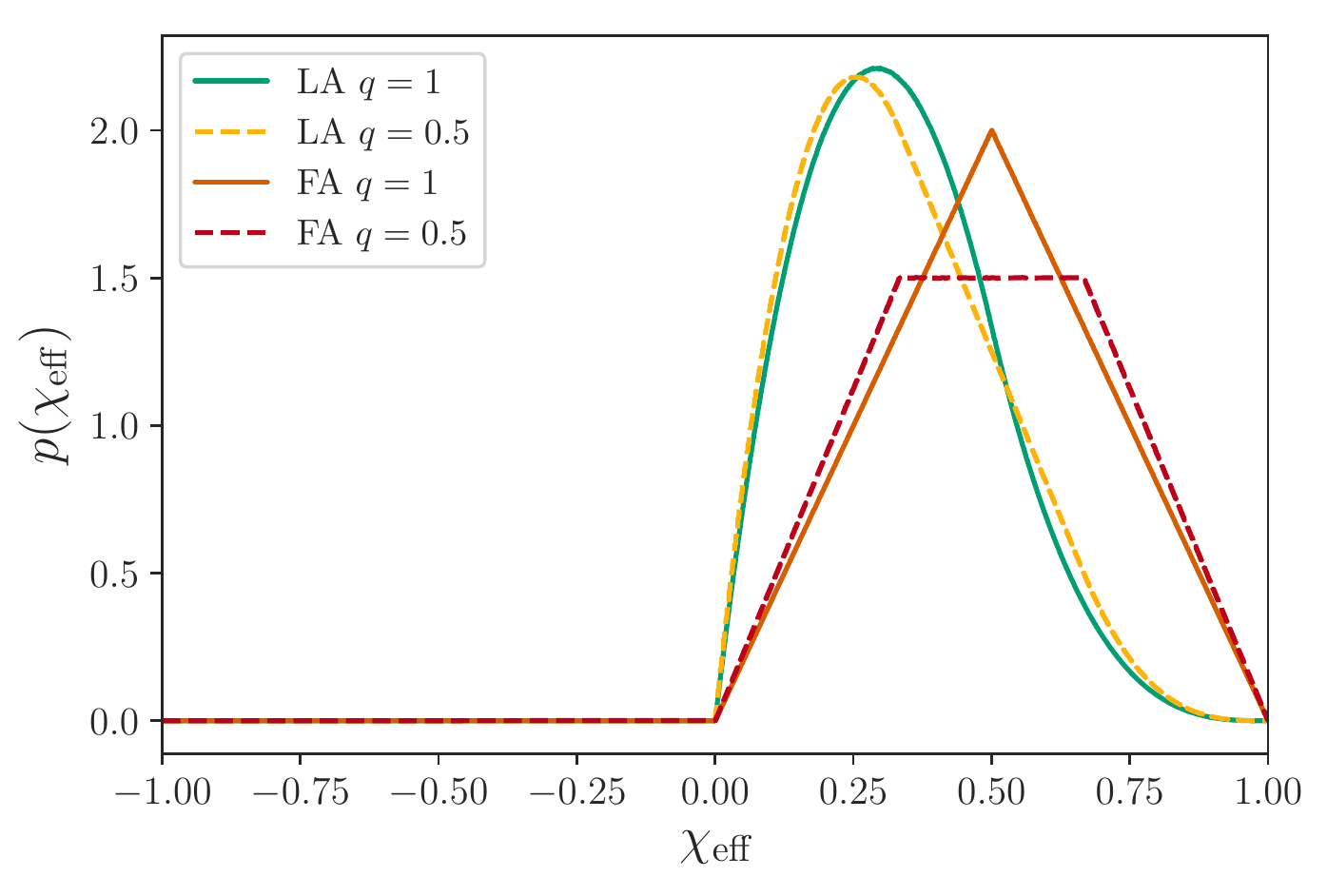}
    \includegraphics[width= .48\textwidth]{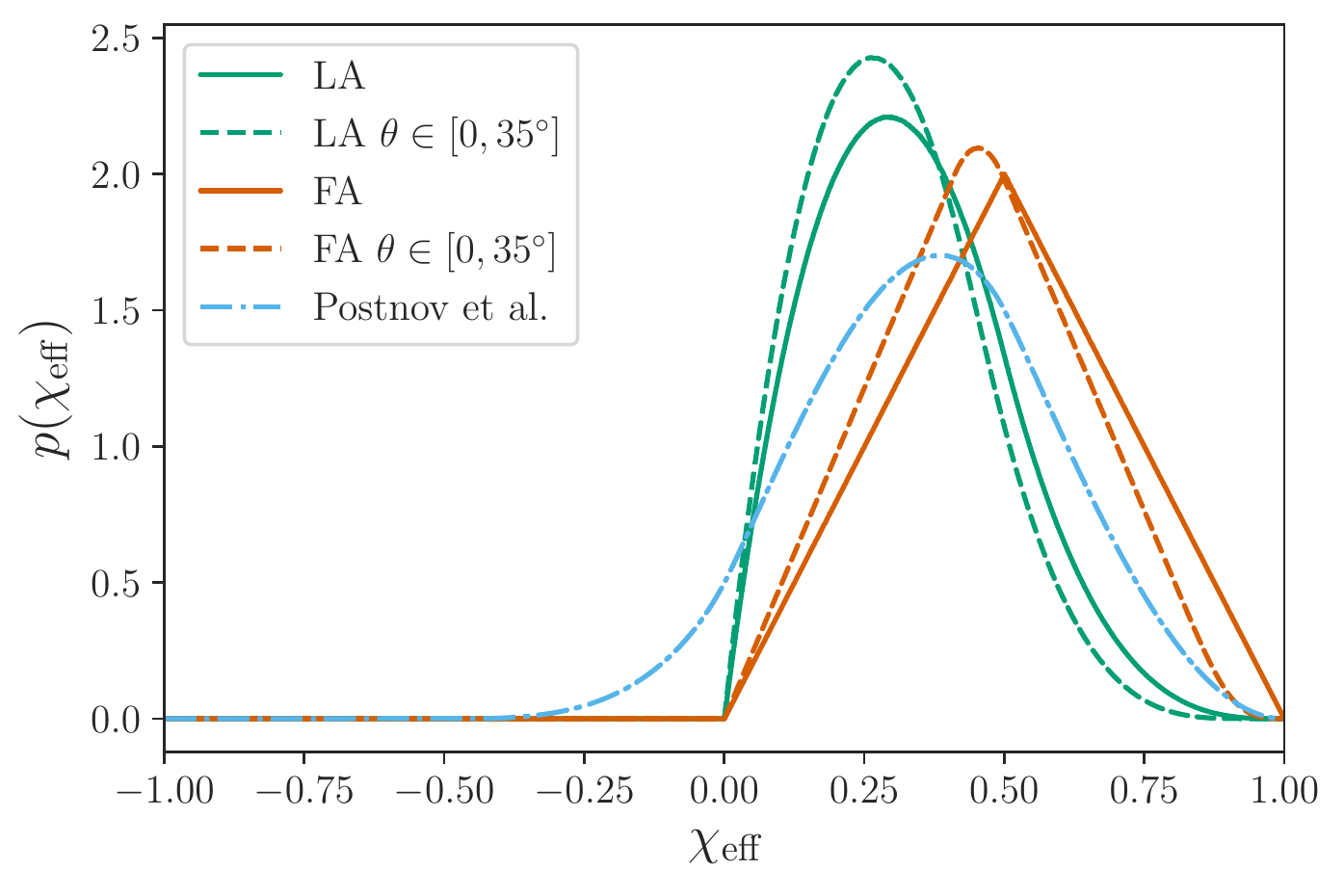}
    \caption{Distribution of $\chieff$ for different mass ratios $q$ for the isotropic (\textit{top left}) and aligned (\textit{top right}) models with $q=1$ and $q=0.5$. \textit{Bottom}: the distribution of $\chieff$ when the assumption of perfect alignment is relaxed for the aligned models (green and orange) and BBH efrom massive isolated binaries through common envelope evolution with randomly misaligned initial spin (blue dash{-}dotted line) \cite{Postnov:2017nfw}.}
    \label{fig:chieff_Priors_sigma_q_35}
\end{figure}

%%%%%%%%%%%%%%%%%%%%
%PBH
%%%%%%%%%%%%%%%%%%%%
\subsection{PBH}
\label{app:PBH}
In figure \ref{fig:chieff_Priors_sigma_q}, we compare the LIGO prior distribution with the results for the prior distribution for $\chieff$ for our PBH models, for various values of $\sigma=0.1,$ 0.2 and 0.3 (top left) and of the mass ratio $q$ (top right).  By allowing the width to vary we get a sense of how sensitive is the PBH distribution with respect to $\sigma$ is. 

The top right panel, instead, illustrates how the choice of $q$ has a very marginal impact on the functional shape of the predicted prior distribution. Notice that all events reported by LIGO have a median value of the mass ratio q larger than $0.5$ and larger than $0.2$ at $90\%$ credible interval (figure 5 on ref. \cite{LIGOScientific:2018mvr}), which is why we focus on relatively large mass ratios, $q=0.2$, 0.5 and 1.0.

Finally, the bottom panel shows our benchmark $q=1$ and $\sigma=0.27$ PBH prior distribution with what is predicted in the models of ref.~\cite{DeLuca:2019buf}. Notice that as illustrated in Fig.~5 of ref.~\cite{DeLuca:2019buf}, the relevant value for $\nu\simeq6$ for masses in the 10 $M_\odot$ range, while the choice of $\gamma$ reflects the prediction for a nearly flat power spectrum ($\gamma=0.88$) and a slightly smaller value, which could result from e.g. a broader power spectrum (see e.g. their Fig.~6, left panel).

%%%%%%%%%%%%%%%%%%%%
%Spin Models
%%%%%%%%%%%%%%%%%%%%
\subsection{Benchmark spin models}
\label{app:Spin_Models}

Here we study the systematic dependence of the prior distribution for $\chieff$ on the values of the mass ratio $q$ (Fig.~\ref{fig:chieff_Priors_sigma_q_35}, top panels) and on the assumption of perfect alignment (bottom panel).

The top left panel shows how the LI, FI and HI prior distributions change when switching to our benchmark value of $q=1$ (equal mass ratio) to $q=0.5$. The top right panel does the same for the aligned models LA and FA. It is clear that the changes for the isotropic models is very small, while for the aligned models there is a slight but noticeable effect near the peak of the distribution for the FA model, and a slight shift in the peak location for the LA model. For the aligned models, as the mass ratio starts to get more extreme, the distribution for $\chieff$ begins to resemble the distribution for the spin magnitude. 

The bottom panel of Fig.~\ref{fig:chieff_Priors_sigma_q_35} shows how the prior distributions for the LA and FA models change when the assumption of perfect BH spin-orbit is relaxed. If the BBH is formed via classical isolated binary evolution with an initial perfect alignment, large values for the misalignment angle are required to be consistent with the GW151226 event \cite{OShaughnessy:2017eks}, but the natal kicks necessary to explain this large misalignments usually exceed the typical values for binary evolution models. Therefore, the distribution of the tilt angle is taken to be flat from $0$ up to $35^{\circ}$, and thus the maximal angle is within the region suggested by ref.{~}\cite{OShaughnessy:2017eks}. The figure shows that the prior distribution is hardly affected at all, with the only noticeable change being a shift of the peak distribution to slightly lower values of $\chieff$. 

Finally, we consider the scenario of binary BH evolution from isolated massive binary systems studied in ref. \cite{Postnov:2017nfw}; in this scenario $\chieff$ can take negative values because of effects such as initially misaligned binary components spins; while other effects such as mass exchange, tidal interaction, and common envelope evolution tend to align the spins, the resulting distribution for $\chieff$ was found to range between -0.2 and 0,8, thus encompassing negative values (see their Figure 11). To mimic this physical situation, here we assumed (following what shown in fig.~8 of ref.~\cite{Postnov:2017nfw}) that one black hole has a spin orientation that is isotropic compared to the orbital angular momentum, and has low intrinsic spin magnitude distribution; for the second black hole, we assumed perfect alignment, and a flat intrinsic spin magnitude distribution. The resulting $\chieff$ distribution is shown in Fig.~\ref{fig:chieff_Priors_sigma_q_35} with a dot-dashed light-blue line; our results reflect qualitatively the range for $\chieff$ found in ref.~\cite{Postnov:2017nfw}. We calculated the resulting odds ratio for this setup, and found that compared to the benchmark LI model, the odd ratio is of -6.04, thus comparable to what we found for the LA model.

\bibliographystyle{JHEP}
\bibliography{references}

\providecommand{\href}[2]{#2}\begingroup\raggedright\begin{thebibliography}{10}

\bibitem{Abbott:2016blz}
{\scshape LIGO Scientific, Virgo} collaboration, B.~P. Abbott et~al.,
  \emph{{Observation of Gravitational Waves from a Binary Black Hole Merger}},
  \href{https://doi.org/10.1103/PhysRevLett.116.061102}{\emph{Phys. Rev. Lett.}
  {\bfseries 116} (2016) 061102}
  [\href{https://arxiv.org/abs/1602.03837}{{\ttfamily 1602.03837}}].

\bibitem{LIGOScientific:2018jsj}
{\scshape LIGO Scientific, Virgo} collaboration, B.~P. Abbott et~al.,
  \emph{{Binary Black Hole Population Properties Inferred from the First and
  Second Observing Runs of Advanced LIGO and Advanced Virgo}},
  \href{https://arxiv.org/abs/1811.12940}{{\ttfamily 1811.12940}}.

\bibitem{Bird:2016dcv}
S.~Bird, I.~Cholis, J.~B. Muñoz, Y.~Ali-Haïmoud, M.~Kamionkowski, E.~D.
  Kovetz et~al., \emph{{Did LIGO detect dark matter?}},
  \href{https://doi.org/10.1103/PhysRevLett.116.201301}{\emph{Phys. Rev. Lett.}
  {\bfseries 116} (2016) 201301}
  [\href{https://arxiv.org/abs/1603.00464}{{\ttfamily 1603.00464}}].

\bibitem{Blinnikov:2016bxu}
S.~Blinnikov, A.~Dolgov, N.~K. Porayko and K.~Postnov, \emph{{Solving puzzles
  of GW150914 by primordial black holes}},
  \href{https://doi.org/10.1088/1475-7516/2016/11/036}{\emph{JCAP} {\bfseries
  1611} (2016) 036} [\href{https://arxiv.org/abs/1611.00541}{{\ttfamily
  1611.00541}}].

\bibitem{Carr:2009jm}
B.~J. Carr, K.~Kohri, Y.~Sendouda and J.~Yokoyama, \emph{{New cosmological
  constraints on primordial black holes}},
  \href{https://doi.org/10.1103/PhysRevD.81.104019}{\emph{Phys. Rev.}
  {\bfseries D81} (2010) 104019}
  [\href{https://arxiv.org/abs/0912.5297}{{\ttfamily 0912.5297}}].

\bibitem{Carr:2016drx}
B.~Carr, F.~Kuhnel and M.~Sandstad, \emph{{Primordial Black Holes as Dark
  Matter}}, \href{https://doi.org/10.1103/PhysRevD.94.083504}{\emph{Phys. Rev.}
  {\bfseries D94} (2016) 083504}
  [\href{https://arxiv.org/abs/1607.06077}{{\ttfamily 1607.06077}}].

\bibitem{Poulin:2017bwe}
V.~Poulin, P.~D. Serpico, F.~Calore, S.~Clesse and K.~Kohri, \emph{{CMB bounds
  on disk-accreting massive primordial black holes}},
  \href{https://doi.org/10.1103/PhysRevD.96.083524}{\emph{Phys. Rev.}
  {\bfseries D96} (2017) 083524}
  [\href{https://arxiv.org/abs/1707.04206}{{\ttfamily 1707.04206}}].

\bibitem{Nakama:2017xvq}
T.~Nakama, B.~Carr and J.~Silk, \emph{{Limits on primordial black holes from
  $\mu$ distortions in cosmic microwave background}},
  \href{https://doi.org/10.1103/PhysRevD.97.043525}{\emph{Phys. Rev.}
  {\bfseries D97} (2018) 043525}
  [\href{https://arxiv.org/abs/1710.06945}{{\ttfamily 1710.06945}}].

\bibitem{Ali-Haimoud:2016mbv}
Y.~Ali-Haïmoud and M.~Kamionkowski, \emph{{Cosmic microwave background limits
  on accreting primordial black holes}},
  \href{https://doi.org/10.1103/PhysRevD.95.043534}{\emph{Phys. Rev.}
  {\bfseries D95} (2017) 043534}
  [\href{https://arxiv.org/abs/1612.05644}{{\ttfamily 1612.05644}}].

\bibitem{Brandt:2016aco}
T.~D. Brandt, \emph{{Constraints on MACHO Dark Matter from Compact Stellar
  Systems in Ultra-Faint Dwarf Galaxies}},
  \href{https://doi.org/10.3847/2041-8205/824/2/L31}{\emph{Astrophys. J.}
  {\bfseries 824} (2016) L31}
  [\href{https://arxiv.org/abs/1605.03665}{{\ttfamily 1605.03665}}].

\bibitem{Koushiappas:2017chw}
S.~M. Koushiappas and A.~Loeb, \emph{{Dynamics of Dwarf Galaxies Disfavor
  Stellar-Mass Black Holes as Dark Matter}},
  \href{https://doi.org/10.1103/PhysRevLett.119.041102}{\emph{Phys. Rev. Lett.}
  {\bfseries 119} (2017) 041102}
  [\href{https://arxiv.org/abs/1704.01668}{{\ttfamily 1704.01668}}].

\bibitem{Li:2016utv}
{\scshape DES} collaboration, T.~S. Li et~al., \emph{{Farthest Neighbor: The
  Distant Milky Way Satellite Eridanus II}},
  \href{https://doi.org/10.3847/1538-4357/aa6113}{\emph{Astrophys. J.}
  {\bfseries 838} (2017) 8} [\href{https://arxiv.org/abs/1611.05052}{{\ttfamily
  1611.05052}}].

\bibitem{Zumalacarregui:2017qqd}
M.~Zumalacarregui and U.~Seljak, \emph{{Limits on stellar-mass compact objects
  as dark matter from gravitational lensing of type Ia supernovae}},
  \href{https://doi.org/10.1103/PhysRevLett.121.141101}{\emph{Phys. Rev. Lett.}
  {\bfseries 121} (2018) 141101}
  [\href{https://arxiv.org/abs/1712.02240}{{\ttfamily 1712.02240}}].

\bibitem{Garcia-Bellido:2017imq}
J.~Garcia-Bellido, S.~Clesse and P.~Fleury, \emph{{Primordial black holes
  survive SN lensing constraints}},
  \href{https://doi.org/10.1016/j.dark.2018.04.005}{\emph{Phys. Dark Univ.}
  {\bfseries 20} (2018) 95} [\href{https://arxiv.org/abs/1712.06574}{{\ttfamily
  1712.06574}}].

\bibitem{Harada:2017fjm}
T.~Harada, C.-M. Yoo, K.~Kohri and K.-I. Nakao, \emph{{Spins of primordial
  black holes formed in the matter-dominated phase of the Universe}},
  \href{https://doi.org/10.1103/PhysRevD.99.069904,
  10.1103/PhysRevD.96.083517}{\emph{Phys. Rev.} {\bfseries D96} (2017) 083517}
  [\href{https://arxiv.org/abs/1707.03595}{{\ttfamily 1707.03595}}].

\bibitem{DEramo:2017gpl}
F.~D'Eramo, N.~Fernandez and S.~Profumo, \emph{{When the Universe Expands Too
  Fast: Relentless Dark Matter}},
  \href{https://doi.org/10.1088/1475-7516/2017/05/012}{\emph{JCAP} {\bfseries
  1705} (2017) 012} [\href{https://arxiv.org/abs/1703.04793}{{\ttfamily
  1703.04793}}].

\bibitem{DEramo:2017ecx}
F.~D'Eramo, N.~Fernandez and S.~Profumo, \emph{{Dark Matter Freeze-in
  Production in Fast-Expanding Universes}},
  \href{https://doi.org/10.1088/1475-7516/2018/02/046}{\emph{JCAP} {\bfseries
  1802} (2018) 046} [\href{https://arxiv.org/abs/1712.07453}{{\ttfamily
  1712.07453}}].

\bibitem{Abbott:2017vtc}
{\scshape VIRGO, LIGO Scientific} collaboration, B.~P. Abbott et~al.,
  \emph{{GW170104: Observation of a 50-Solar-Mass Binary Black Hole Coalescence
  at Redshift 0.2}}, \href{https://doi.org/10.1103/PhysRevLett.118.221101,
  10.1103/PhysRevLett.121.129901}{\emph{Phys. Rev. Lett.} {\bfseries 118}
  (2017) 221101} [\href{https://arxiv.org/abs/1706.01812}{{\ttfamily
  1706.01812}}].

\bibitem{Abbott:2017gyy}
{\scshape Virgo, LIGO Scientific} collaboration, B.~P. Abbott et~al.,
  \emph{{GW170608: Observation of a 19-solar-mass Binary Black Hole
  Coalescence}},
  \href{https://doi.org/10.3847/2041-8213/aa9f0c}{\emph{Astrophys. J.}
  {\bfseries 851} (2017) L35}
  [\href{https://arxiv.org/abs/1711.05578}{{\ttfamily 1711.05578}}].

\bibitem{Abbott:2017oio}
{\scshape Virgo, LIGO Scientific} collaboration, B.~P. Abbott et~al.,
  \emph{{GW170814: A Three-Detector Observation of Gravitational Waves from a
  Binary Black Hole Coalescence}},
  \href{https://doi.org/10.1103/PhysRevLett.119.141101}{\emph{Phys. Rev. Lett.}
  {\bfseries 119} (2017) 141101}
  [\href{https://arxiv.org/abs/1709.09660}{{\ttfamily 1709.09660}}].

\bibitem{TheLIGOScientific:2016pea}
{\scshape Virgo, LIGO Scientific} collaboration, B.~P. Abbott et~al.,
  \emph{{Binary Black Hole Mergers in the first Advanced LIGO Observing Run}},
  \href{https://doi.org/10.1103/PhysRevX.6.041015,
  10.1103/PhysRevX.8.039903}{\emph{Phys. Rev.} {\bfseries X6} (2016) 041015}
  [\href{https://arxiv.org/abs/1606.04856}{{\ttfamily 1606.04856}}].

\bibitem{Farr:2017uvj}
W.~M. Farr, S.~Stevenson, M.~Coleman~Miller, I.~Mandel, B.~Farr and A.~Vecchio,
  \emph{{Distinguishing Spin-Aligned and Isotropic Black Hole Populations With
  Gravitational Waves}},
  \href{https://doi.org/10.1038/nature23453}{\emph{Nature} {\bfseries 548}
  (2017) 426} [\href{https://arxiv.org/abs/1706.01385}{{\ttfamily
  1706.01385}}].

\bibitem{Belczynski:2017gds}
K.~Belczynski et~al., \emph{{The origin of low spin of black holes in
  LIGO/Virgo mergers}},  \href{https://arxiv.org/abs/1706.07053}{{\ttfamily
  1706.07053}}.

\bibitem{Vitale:2017cfs}
S.~Vitale, D.~Gerosa, C.-J. Haster, K.~Chatziioannou and A.~Zimmerman,
  \emph{{Impact of Bayesian Priors on the Characterization of Binary Black Hole
  Coalescences}},
  \href{https://doi.org/10.1103/PhysRevLett.119.251103}{\emph{Phys. Rev. Lett.}
  {\bfseries 119} (2017) 251103}
  [\href{https://arxiv.org/abs/1707.04637}{{\ttfamily 1707.04637}}].

\bibitem{Farr:2017gtv}
B.~Farr, D.~E. Holz and W.~M. Farr, \emph{{Using Spin to Understand the
  Formation of LIGO and Virgo’s Black Holes}},
  \href{https://doi.org/10.3847/2041-8213/aaaa64}{\emph{Astrophys. J.}
  {\bfseries 854} (2018) L9}
  [\href{https://arxiv.org/abs/1709.07896}{{\ttfamily 1709.07896}}].

\bibitem{Vitale:2015tea}
S.~Vitale, R.~Lynch, R.~Sturani and P.~Graff, \emph{{Use of gravitational waves
  to probe the formation channels of compact binaries}},
  \href{https://doi.org/10.1088/1361-6382/aa552e}{\emph{Class. Quant. Grav.}
  {\bfseries 34} (2017) 03LT01}
  [\href{https://arxiv.org/abs/1503.04307}{{\ttfamily 1503.04307}}].

\bibitem{Stevenson:2017dlk}
S.~Stevenson, C.~P.~L. Berry and I.~Mandel, \emph{{Hierarchical analysis of
  gravitational-wave measurements of binary black hole spin–orbit
  misalignments}}, \href{https://doi.org/10.1093/mnras/stx1764}{\emph{Mon. Not.
  Roy. Astron. Soc.} {\bfseries 471} (2017) 2801}
  [\href{https://arxiv.org/abs/1703.06873}{{\ttfamily 1703.06873}}].

\bibitem{Wysocki:2018mpo}
D.~Wysocki, J.~Lange and R.~O.~'shaughnessy, \emph{{Reconstructing
  phenomenological distributions of compact binaries via gravitational wave
  observations}},  \href{https://arxiv.org/abs/1805.06442}{{\ttfamily
  1805.06442}}.

\bibitem{Chiba:2017rvs}
T.~Chiba and S.~Yokoyama, \emph{{Spin Distribution of Primordial Black Holes}},
  \href{https://doi.org/10.1093/ptep/ptx087}{\emph{PTEP} {\bfseries 2017}
  (2017) 083E01} [\href{https://arxiv.org/abs/1704.06573}{{\ttfamily
  1704.06573}}].

\bibitem{DeLuca:2019buf}
V.~De~Luca, V.~Desjacques, G.~Franciolini, A.~Malhotra and A.~Riotto,
  \emph{{The Initial Spin Probability Distribution of Primordial Black Holes}},
   \href{https://arxiv.org/abs/1903.01179}{{\ttfamily 1903.01179}}.

\bibitem{Mirbabayi:2019uph}
M.~Mirbabayi, A.~Gruzinov and J.~Noreña, \emph{{Spin of Primordial Black
  Holes}},  \href{https://arxiv.org/abs/1901.05963}{{\ttfamily 1901.05963}}.

\bibitem{He:2019cdb}
M.~He and T.~Suyama, \emph{{Formation threshold of rotating primordial black
  holes}},  \href{https://arxiv.org/abs/1906.10987}{{\ttfamily 1906.10987}}.

\bibitem{LIGOScientific:2018mvr}
{\scshape LIGO Scientific, Virgo} collaboration, B.~P. Abbott et~al.,
  \emph{{GWTC-1: A Gravitational-Wave Transient Catalog of Compact Binary
  Mergers Observed by LIGO and Virgo during the First and Second Observing
  Runs}},  \href{https://arxiv.org/abs/1811.12907}{{\ttfamily 1811.12907}}.

\bibitem{Abadie:2010cf}
{\scshape LIGO Scientific, VIRGO} collaboration, J.~Abadie et~al.,
  \emph{{Predictions for the Rates of Compact Binary Coalescences Observable by
  Ground-based Gravitational-wave Detectors}},
  \href{https://doi.org/10.1088/0264-9381/27/17/173001}{\emph{Class. Quant.
  Grav.} {\bfseries 27} (2010) 173001}
  [\href{https://arxiv.org/abs/1003.2480}{{\ttfamily 1003.2480}}].

\bibitem{Lipunov:1996nj}
V.~M. Lipunov, K.~A. Postnov and M.~E. Prokhorov, \emph{{First LIGO events:
  Binary black holes mergings}},
  \href{https://doi.org/10.1016/S1384-1076(97)00007-9}{\emph{New Astron.}
  {\bfseries 2} (1997) 43}
  [\href{https://arxiv.org/abs/astro-ph/9610016}{{\ttfamily
  astro-ph/9610016}}].

\bibitem{Lipunov:1997xm}
V.~M. Lipunov, K.~A. Postnov and M.~E. Prokhorov, \emph{{Black holes and
  gravitational waves: Simultaneous discovery by initial laser
  interferometers}}, {\emph{Submitted to: Astron. Lett.} (1997) }
  [\href{https://arxiv.org/abs/astro-ph/9701134}{{\ttfamily
  astro-ph/9701134}}].

\bibitem{Mandel:2009nx}
I.~Mandel and R.~O'Shaughnessy, \emph{{Compact Binary Coalescences in the Band
  of Ground-based Gravitational-Wave Detectors}},
  \href{https://doi.org/10.1088/0264-9381/27/11/114007}{\emph{Class. Quant.
  Grav.} {\bfseries 27} (2010) 114007}
  [\href{https://arxiv.org/abs/0912.1074}{{\ttfamily 0912.1074}}].

\bibitem{Spera:2017fyx}
M.~Spera and M.~Mapelli, \emph{{Very massive stars, pair-instability supernovae
  and intermediate-mass black holes with the SEVN code}},
  \href{https://doi.org/10.1093/mnras/stx1576}{\emph{Mon. Not. Roy. Astron.
  Soc.} {\bfseries 470} (2017) 4739}
  [\href{https://arxiv.org/abs/1706.06109}{{\ttfamily 1706.06109}}].

\bibitem{Kimball:2019mfs}
C.~Kimball, C.~P.~L. Berry and V.~Kalogera, \emph{{What GW170729's exceptional
  mass and spin tells us about its family tree}},
  \href{https://arxiv.org/abs/1903.07813}{{\ttfamily 1903.07813}}.

\bibitem{Kruckow:2018slo}
M.~U. Kruckow, T.~M. Tauris, N.~Langer, M.~Kramer and R.~G. Izzard,
  \emph{{Progenitors of gravitational wave mergers: Binary evolution with the
  stellar grid-based code ComBinE}},
  \href{https://arxiv.org/abs/1801.05433}{{\ttfamily 1801.05433}}.

\bibitem{Fishbach:2017dwv}
M.~Fishbach, D.~E. Holz and B.~Farr, \emph{{Are LIGO's Black Holes Made From
  Smaller Black Holes?}},
  \href{https://doi.org/10.3847/2041-8213/aa7045}{\emph{Astrophys. J.}
  {\bfseries 840} (2017) L24}
  [\href{https://arxiv.org/abs/1703.06869}{{\ttfamily 1703.06869}}].

\bibitem{Vitale:2016aso}
S.~Vitale, \emph{{Three observational differences for binary black holes
  detections with second and third generation gravitational-wave detectors}},
  \href{https://doi.org/10.1103/PhysRevD.94.121501}{\emph{Phys. Rev.}
  {\bfseries D94} (2016) 121501}
  [\href{https://arxiv.org/abs/1610.06914}{{\ttfamily 1610.06914}}].

\bibitem{Heger:2001cd}
A.~Heger and S.~E. Woosley, \emph{{The nucleosynthetic signature of population
  III}}, \href{https://doi.org/10.1086/338487}{\emph{Astrophys. J.} {\bfseries
  567} (2002) 532} [\href{https://arxiv.org/abs/astro-ph/0107037}{{\ttfamily
  astro-ph/0107037}}].

\bibitem{Belczynski:2016jno}
K.~Belczynski et~al., \emph{{The Effect of Pair-Instability Mass Loss on Black
  Hole Mergers}},
  \href{https://doi.org/10.1051/0004-6361/201628980}{\emph{Astron. Astrophys.}
  {\bfseries 594} (2016) A97}
  [\href{https://arxiv.org/abs/1607.03116}{{\ttfamily 1607.03116}}].

\bibitem{Fishbach:2017zga}
M.~Fishbach and D.~E. Holz, \emph{{Where Are LIGO’s Big Black Holes?}},
  \href{https://doi.org/10.3847/2041-8213/aa9bf6}{\emph{Astrophys. J.}
  {\bfseries 851} (2017) L25}
  [\href{https://arxiv.org/abs/1709.08584}{{\ttfamily 1709.08584}}].

\bibitem{Talbot:2017yur}
C.~Talbot and E.~Thrane, \emph{{Determining the population properties of
  spinning black holes}},
  \href{https://doi.org/10.1103/PhysRevD.96.023012}{\emph{Phys. Rev.}
  {\bfseries D96} (2017) 023012}
  [\href{https://arxiv.org/abs/1704.08370}{{\ttfamily 1704.08370}}].

\bibitem{Roulet:2018jbe}
J.~Roulet and M.~Zaldarriaga, \emph{{Constraints on Binary Black Hole
  Populations from LIGO-Virgo Detections}},
  \href{https://doi.org/10.1093/mnras/stz226}{\emph{Mon. Not. Roy. Astron.
  Soc.} {\bfseries 484} (2019) 4216}
  [\href{https://arxiv.org/abs/1806.10610}{{\ttfamily 1806.10610}}].

\bibitem{Bai:2018shq}
Y.~Bai, V.~Barger and S.~Lu, \emph{{Measuring the Black Hole Mass Spectrum from
  Redshifts of aLIGO Binary Merger Events}},
  \href{https://arxiv.org/abs/1802.04909}{{\ttfamily 1802.04909}}.

\bibitem{McClintock:2013vwa}
J.~E. McClintock, R.~Narayan and J.~F. Steiner, \emph{{Black Hole Spin via
  Continuum Fitting and the Role of Spin in Powering Transient Jets}},
  \href{https://doi.org/10.1007/s11214-013-0003-9}{\emph{Space Sci. Rev.}
  {\bfseries 183} (2014) 295}
  [\href{https://arxiv.org/abs/1303.1583}{{\ttfamily 1303.1583}}].

\bibitem{Gou:2011nq}
L.~Gou, J.~E. McClintock, M.~J. Reid, J.~A. Orosz, J.~F. Steiner, R.~Narayan
  et~al., \emph{{The Extreme Spin of the Black Hole in Cygnus X-1}},
  \href{https://doi.org/10.1088/0004-637X/742/2/85}{\emph{Astrophys. J.}
  {\bfseries 742} (2011) 85} [\href{https://arxiv.org/abs/1106.3690}{{\ttfamily
  1106.3690}}].

\bibitem{Thrane:2018qnx}
E.~Thrane and C.~Talbot, \emph{{An introduction to Bayesian inference in
  gravitational-wave astronomy: parameter estimation, model selection, and
  hierarchical models}}, \href{https://doi.org/10.1017/pasa.2019.2}{\emph{Publ.
  Astron. Soc. Austral.} {\bfseries 36} (2019) 10}
  [\href{https://arxiv.org/abs/1809.02293}{{\ttfamily 1809.02293}}].

\bibitem{Ng:2018neg}
K.~K.~Y. Ng, S.~Vitale, A.~Zimmerman, K.~Chatziioannou, D.~Gerosa and C.-J.
  Haster, \emph{{Gravitational-wave astrophysics with effective-spin
  measurements: asymmetries and selection biases}},
  \href{https://doi.org/10.1103/PhysRevD.98.083007}{\emph{Phys. Rev.}
  {\bfseries D98} (2018) 083007}
  [\href{https://arxiv.org/abs/1805.03046}{{\ttfamily 1805.03046}}].

\bibitem{Campanelli:2006uy}
M.~Campanelli, C.~O. Lousto and Y.~Zlochower, \emph{{Spinning-black-hole
  binaries: The orbital hang up}},
  \href{https://doi.org/10.1103/PhysRevD.74.041501}{\emph{Phys. Rev.}
  {\bfseries D74} (2006) 041501}
  [\href{https://arxiv.org/abs/gr-qc/0604012}{{\ttfamily gr-qc/0604012}}].

\bibitem{Clesse:2017bsw}
S.~Clesse and J.~García-Bellido, \emph{{Seven Hints for Primordial Black Hole
  Dark Matter}}, \href{https://doi.org/10.1016/j.dark.2018.08.004}{\emph{Phys.
  Dark Univ.} {\bfseries 22} (2018) 137}
  [\href{https://arxiv.org/abs/1711.10458}{{\ttfamily 1711.10458}}].

\bibitem{Tiwari:2018qch}
V.~Tiwari, S.~Fairhurst and M.~Hannam, \emph{{Constraining black-hole spins
  with gravitational wave observations}},
  \href{https://doi.org/10.3847/1538-4357/aae8df}{\emph{Astrophys. J.}
  {\bfseries 868} (2018) 140}
  [\href{https://arxiv.org/abs/1809.01401}{{\ttfamily 1809.01401}}].

\bibitem{1993MNRAS.260..675T}
A.~V. Tutukov and L.~R. YungelSon, \emph{{The merger rate of neutron star and
  black hole binaries}},
  \href{https://doi.org/10.1093/mnras/260.3.675}{\emph{Mon. Not. Roy. Astron.
  Soc.} {\bfseries 260} (1993) 675}.

\bibitem{Miller:2014aaa}
M.~C. Miller and J.~M. Miller, \emph{{The Masses and Spins of Neutron Stars and
  Stellar-Mass Black Holes}},
  \href{https://doi.org/10.1016/j.physrep.2014.09.003}{\emph{Phys. Rept.}
  {\bfseries 548} (2014) 1} [\href{https://arxiv.org/abs/1408.4145}{{\ttfamily
  1408.4145}}].

\bibitem{Gerosa:2018wbw}
D.~Gerosa, E.~Berti, R.~O'Shaughnessy, K.~Belczynski, M.~Kesden, D.~Wysocki
  et~al., \emph{{Spin orientations of merging black holes formed from the
  evolution of stellar binaries}},
  \href{https://doi.org/10.1103/PhysRevD.98.084036}{\emph{Phys. Rev.}
  {\bfseries D98} (2018) 084036}
  [\href{https://arxiv.org/abs/1808.02491}{{\ttfamily 1808.02491}}].

\bibitem{2010ApJ...719L..79F}
T.~{Fragos}, M.~{Tremmel}, E.~{Rantsiou} and K.~{Belczynski}, \emph{{Black Hole
  Spin-Orbit Misalignment in Galactic X-ray Binaries}},
  \href{https://doi.org/10.1088/2041-8205/719/1/L79}{\emph{Astrophys. J.}
  {\bfseries 719} (2010) L79}
  [\href{https://arxiv.org/abs/1001.1107}{{\ttfamily 1001.1107}}].

\bibitem{Wysocki:2017isg}
D.~Wysocki, D.~Gerosa, R.~O'Shaughnessy, K.~Belczynski, W.~Gladysz, E.~Berti
  et~al., \emph{{Explaining LIGO’s observations via isolated binary evolution
  with natal kicks}},
  \href{https://doi.org/10.1103/PhysRevD.97.043014}{\emph{Phys. Rev.}
  {\bfseries D97} (2018) 043014}
  [\href{https://arxiv.org/abs/1709.01943}{{\ttfamily 1709.01943}}].

\bibitem{TheLIGOScientific:2016htt}
{\scshape LIGO Scientific, Virgo} collaboration, B.~P. Abbott et~al.,
  \emph{{Astrophysical Implications of the Binary Black-Hole Merger GW150914}},
  \href{https://doi.org/10.3847/2041-8205/818/2/L22}{\emph{Astrophys. J.}
  {\bfseries 818} (2016) L22}
  [\href{https://arxiv.org/abs/1602.03846}{{\ttfamily 1602.03846}}].

\bibitem{Marchant:2016wow}
P.~Marchant, N.~Langer, P.~Podsiadlowski, T.~M. Tauris and T.~J. Moriya,
  \emph{{A new route towards merging massive black holes}},
  \href{https://doi.org/10.1051/0004-6361/201628133}{\emph{Astron. Astrophys.}
  {\bfseries 588} (2016) A50}
  [\href{https://arxiv.org/abs/1601.03718}{{\ttfamily 1601.03718}}].

\bibitem{PortegiesZwart:1999nm}
S.~F. Portegies~Zwart and S.~McMillan, \emph{{Black hole mergers in the
  universe}}, \href{https://doi.org/10.1086/312422}{\emph{Astrophys. J.}
  {\bfseries 528} (2000) L17}
  [\href{https://arxiv.org/abs/astro-ph/9910061}{{\ttfamily
  astro-ph/9910061}}].

\bibitem{Rodriguez:2015oxa}
C.~L. Rodriguez, M.~Morscher, B.~Pattabiraman, S.~Chatterjee, C.-J. Haster and
  F.~A. Rasio, \emph{{Binary Black Hole Mergers from Globular Clusters:
  Implications for Advanced LIGO}},
  \href{https://doi.org/10.1103/PhysRevLett.116.029901,
  10.1103/PhysRevLett.115.051101}{\emph{Phys. Rev. Lett.} {\bfseries 115}
  (2015) 051101} [\href{https://arxiv.org/abs/1505.00792}{{\ttfamily
  1505.00792}}].

\bibitem{Stone:2016wzz}
N.~C. Stone, B.~D. Metzger and Z.~Haiman, \emph{{Assisted inspirals of stellar
  mass black holes embedded in AGN discs: solving the ‘final au problem’}},
  \href{https://doi.org/10.1093/mnras/stw2260}{\emph{Mon. Not. Roy. Astron.
  Soc.} {\bfseries 464} (2017) 946}
  [\href{https://arxiv.org/abs/1602.04226}{{\ttfamily 1602.04226}}].

\bibitem{Sigurdsson:1993zrm}
S.~Sigurdsson and L.~Hernquist, \emph{{Primordial black holes in globular
  clusters}}, \href{https://doi.org/10.1038/364423a0}{\emph{Nature} {\bfseries
  364} (1993) 423}.

\bibitem{Postnov:2019tmw}
K.~Postnov and N.~Mitichkin, \emph{{Spins of primordial binary black holes
  before coalescence}},  \href{https://arxiv.org/abs/1904.00570}{{\ttfamily
  1904.00570}}.

\bibitem{Zackay:2019tzo}
B.~Zackay, T.~Venumadhav, L.~Dai, J.~Roulet and M.~Zaldarriaga, \emph{{A Highly
  Spinning and Aligned Binary Black Hole Merger in the Advanced LIGO First
  Observing Run}},  \href{https://arxiv.org/abs/1902.10331}{{\ttfamily
  1902.10331}}.

\bibitem{Venumadhav:2019lyq}
T.~Venumadhav, B.~Zackay, J.~Roulet, L.~Dai and M.~Zaldarriaga, \emph{{New
  Binary Black Hole Mergers in the Second Observing Run of Advanced LIGO and
  Advanced Virgo}},  \href{https://arxiv.org/abs/1904.07214}{{\ttfamily
  1904.07214}}.

\bibitem{Postnov:2017nfw}
K.~Postnov and A.~Kuranov, \emph{{Black hole spins in coalescing binary black
  holes}}, \href{https://doi.org/10.1093/mnras/sty3313}{\emph{Mon. Not. Roy.
  Astron. Soc.} {\bfseries 483} (2019) 3288}
  [\href{https://arxiv.org/abs/1706.00369}{{\ttfamily 1706.00369}}].

\bibitem{OShaughnessy:2017eks}
R.~O'Shaughnessy, D.~Gerosa and D.~Wysocki, \emph{{Inferences about supernova
  physics from gravitational-wave measurements: GW151226 spin misalignment as
  an indicator of strong black-hole natal kicks}},
  \href{https://doi.org/10.1103/PhysRevLett.119.011101}{\emph{Phys. Rev. Lett.}
  {\bfseries 119} (2017) 011101}
  [\href{https://arxiv.org/abs/1704.03879}{{\ttfamily 1704.03879}}].

\end{thebibliography}\endgroup



\providecommand{\href}[2]{#2}\begingroup\raggedright\endgroup

\end{document}